\documentclass[twocolumn,10pt]{IEEEtran}
\usepackage{stackengine}
\usepackage{amsmath}
\usepackage{amsfonts}
\usepackage{mathrsfs}
\usepackage{dsfont}
\usepackage{eucal}
\usepackage{pifont}
\usepackage{float}
\usepackage{amssymb}
\usepackage{verbatim}
\usepackage{upgreek}
\usepackage{color}
\usepackage[final]{graphicx}
\usepackage{subfigure}
\usepackage{epsfig}
\usepackage{booktabs}
\usepackage{bm}
\usepackage{setspace}
\usepackage{float} 
\usepackage{colortbl}
\usepackage{xcolor} 
\usepackage{multirow}
\usepackage{rotating}
\usepackage{cite}
\usepackage{lipsum}
\usepackage{mathtools}
\usepackage{cuted}

\usepackage[colorlinks]{hyperref}

\begin{document}

\title{Geostationary Real-Time 3D Polarimetric RADAR Imaging by Orbital Angular Momentum Interferometry and Multi-Chromatic Analysis}

\author{\IEEEauthorblockN{Filippo Biondi\IEEEauthorrefmark{1}, Pia Addabbo\IEEEauthorrefmark{2}, Carmine Clemente\IEEEauthorrefmark{3}, Danilo Orlando\IEEEauthorrefmark{4}, and Fabrizio Tamburini\IEEEauthorrefmark{5}
}

\IEEEauthorblockA{\IEEEauthorrefmark{1} Italian Ministry of Defense, Italy. E-mail: biopippoo@gmail.com}

\IEEEauthorblockA{\IEEEauthorrefmark{2} Universit\'a ``Giustino Fortunato'', 82100 Benevento, Italy. E-mail: p.addabbo@unifortunato.eu}

\IEEEauthorblockA{\IEEEauthorrefmark{3} University of Strathclyde, 204 George Street, G1 1XW, Glasgow, Scotland, UK. E-mail: carmine.clemente@strath.ac.uk}\\

\IEEEauthorblockA{\IEEEauthorrefmark{4} Universit\`a degli Studi ``Niccol\`o Cusano'', 00166 Roma, Italy. E-mail: danilo.orlando@unicusano.it} 

\IEEEauthorblockA{\IEEEauthorrefmark{5} ZKM Karlsruhe, 76135 Karlsruhe, Germany. E-mail: fabrizio.tamburini@gmail.com} 

\thanks{Manuscript received December 1, 2012; revised August 26, 2015. 
Corresponding author: Filippo Biondi (email: biopippoo@gmail.com).}}

\maketitle

\begin{abstract}
We design the proof of concept for high-resolution (HR) real-time (RT), Geosynchronous and Geostationary (Geo) Polarimetric (Pol) using orbital angular momentum (OAM) interferometry -  radio detection and ranging (RADAR) (HR-RT-GeoPolInt-OAM-RADAR) and multi-chromatic analysis (MCA) extended to Tomography (HR-RT-GeoPolInt-OAM-MCA-TomoRADAR). 
We propose this technique to replace or to crucially improve the already existing methods based on the Doppler channel for azimuth imaging.
The OAM interferometry communication channel is generated by two fixed sources distanced by a given spatial baseline and used for range-azimuth synthesis. The frequency channel, instead,  is used to provide information about the altitude. 
Finally, the information encoded in the polarization of the electromagnetic (EM) waves, which is related to the Spin Angular Momentum (SAM), is used to synthesize full-Pol RADAR images, with technological redundancy. 
Here we present the design of a planar vortex antenna, tailored for Geo applications, where the imaging system transmits ''ad-hoc`` structured wave packets using an incremental stepped chirp strategy and with single-mode OAM linearly incremented modulation. 
We assign the resolutions of each dimension to three bands assumed by the EM wave. 
The radial and tangential components received from the HR-RT-GeoPolInt-OAM communication channel backscattered signals are used to focus, through fast-Fourier transform (FFT) techniques, a range-azimuth image that belongs to a single epoch at a given constant frequency. 
Each OAM fast-time RADAR image is separated in frequency by using MCA. 
This procedure is repeated for all the epochs of the entire stepped-frequency chirp. 
Once each two-dimensional image is synthesized, they are co-registered, and the HR-RT-GeoPolInt-OAM-MCA-TomoRADAR slices are focused in altitude by using FFT techniques. Range-azimuth and tomographic resolutions depend on the OAM value and the stepped frequency chirp bandwidths. 
We are confident to get a two-dimensional frame about every 10 seconds and a complete tomographic product every 2 minutes. The ''stop-and-go`` approximation is eliminated, so all targets, even those in motion, can be displayed correctly, without any delocalization or cancellation effects.
\end{abstract}


\section{Introduction}
\IEEEPARstart{E}{Lectromagnetic} waves carry energy and momentum; this latter term is represents two independent conserved quantities, the linear momentum $\mathbf{P}$, related to spatial translations, and the total angular momentum $\mathbf{J}$, related to rotations. 
In particular, the total angular momentum conserved quantity $\mathbf{J}$ can be split in two components $\mathbf{S}$ and $\mathbf{L}$ that cannot be conserved separately during the evolution of the system, unless adopting paraxial approximations. 
The first quantity, $\mathbf{J}$, is the spin angular momentum (SAM) and is related to the polarization of an EM wave, whilst the second quantity, $\mathbf{S}$ is depends on the field intensity and phase spatial distributions and is known as orbital angular momentum (OAM). 
\paragraph{OAM Introduction} According to \cite{bethe1932wechselwirkung,belinfante1939theory} the relationship between SAM and OAM can be naively figured by referring to the model of electron rotation around the nucleus in a classical atom, where the momentum that generated its curved trajectory around the nucleus is equivalent to OAM and the momentum generated by the spin of electrons is equivalent to SAM. 
As theorized by Abraham in $1914$ and Majorana \cite{abraham,esposito2006manoscritto}, Allen et al.  \cite{allen1992orbital} experimentally proved with Laguerre-Gaussian laser light that also photons can carry, together with the spin angular momentum, a well-defined quantity of OAM, expressed in integer multiples $\ell$ of the reduced Planck's constant $\hbar$, and measured  the induced mechanical transferred torque \cite{Torres&Torner2011}. 
This tower of OAM values associated to a single photon is related to the spatial intensity and phase distribution of the OAM beam to which it is associated: each OAM beam is characterized by a number $\ell$ of twists in its azimuthal phase and each photon in the beam carries $\ell \hbar$ orbital angular momentum.
This property is slightly different from the original tower of massive particles that are are antiparticles of themselves hypothesized by Majorana in 1932 and instead present an infinite discrete spectrum of intrinsic spin angular momenta \cite{di1932teoria}.
Involving the total angular momentum $\mathbf{J}$, one instead expects towers of OAM and SAM values from organized structures in a resonant astrophysical plasma that behave as quasiparticles carrying well defined quantities of both OAM and SAM \cite{plasma1,plasma2}.

Recently, OAM research has been extended also to radio waves and more specifically to microwaves \cite{thide2007utilization}, in particular, to radio telecommunication for the possibility of implementing multiple channel access. In \cite{thide2011radio} authors show numerically, that vector antenna arrays can generate radio beams that exhibit spin and OAM, characteristics similar to those of helical Laguerre-Gauss laser beams in paraxial optics. The work announce also the efficient applications on radio astronomy and paving the way for novel wireless communication medium sharing concepts. 

OAM waves pave the way to a new type of multiplexing for telecommunications.
The first radio-electric signals, successfully transmitted and received, were realized in the early 1900s by Nikola Tesla and Guglielmo Marconi, they started the wireless communication revolution \cite{Tesla,Marconi,dunlap1937man} by using the linear momentum carried by the EM waves. Then, thanks to the mathematical formalism used by Ettore Majorana, and to his unpublished research \cite{mignani1974dirac}, before 1930, a great similarity between the wave equation for the electron (or even better for the neutrino) and the Maxwell equations and the photon wavefunction \cite{tamburini2008photon} became evident. 
In \cite{tamburini2012encoding} authors show experimentally, in a public experiment in 2011, that it is possible to use two incoherent beams of radio waves, transmitted on the same frequency but encoded in two different OAM states, to simultaneously transmit two independent radio channels. This novel radio technique allows the implementation of, in principle, an infinite number of channels in a given, fixed frequency bandwidth, even without using polarization, multi-port or dense coding techniques
then confirmed almost at the same time in optical communications \cite{huang2014100}. This solution paves the way for innovative techniques in radio science and entirely new paradigms for radio communication protocols that might offer a solution to the problem of radio-band congestion \cite{union2008itu}. In \cite{tamburini2020measurement} presented the first observational evidence that light propagating near a rotating black hole is twisted in phase and carries OAM. This physical observable allows a direct measurement of the rotation of the black hole.
In the pionieristic works are presented several experiments, first steps in a novel field where the authors characterize circular arrays of patch antennas designed to transmit and receive OAM-EM fields for a short-range experimental communication system  \cite{spinello2015experimental}, or to realize successfully the tripling of the channel capacity of any given polarization state \cite{triple}, communicating with superimposed opposite OAM modes \cite{someda} and finally discussed about the divergence of OAM beams and multiplexing capacity when only the intensity and spatial phase distribution is taken in account \cite{oldoni}, demonstrating the potential applications to communication links based on OAM modes that are currently a hot topic for telecommunications. In \cite{rui2019orbital} authors summarize the methods for the generation and detection of optical OAM, radio OAM, and acoustic OAM. 
In \cite{liu2014orbital} a novel radio detection and ranging (RADAR) imaging technique based on OAM modulation is presented. 
First they generate and distribute the OAM-EM vortex wave using incrementally phased uniform circular arrays and then analyze the factors that affect the phase front distribution. The work can benefit the development of novel information-rich RADAR based on OAM, as well as RADAR target recognition. 
In \cite{sun2018realization} A novel experimental platform is developed to measure the OAM phase distribution when the transmitting receiving antennas operate at different frequencies for multi-input multi-output (MIMO) synthetic aperture RADAR (SAR) applications. In \cite{bu2018novel} an OAM-based MIMO-SAR system is proposed, theorizing high-resolution (HR) wide-swath (HRWS) remote sensing, 3-D imaging, and RADAR-communication integration. The concept of RADAR imaging based on OAM, has the ability of azimuthal resolution without relative motion, has been proposed in \cite{guirong2013electromagnetic} while in \cite{yuan2017radar} authors investigate the phase and intensity distributions of the imaging model showing that the received signal has the form of an inverse discrete Fourier transform with the use of OAM and frequency diversities. A practical concerning OAM-SAR imaging has been included in \cite{bu2018implementation} and a theoretical study of 3D SAR focusing has been considered in \cite{yang2018three}. Concluding, the authors of \cite{bu2019synthetic} applied OAM for interferometric applications in order to retrieve topographic heights from monocular SAR observations. 
\paragraph{Geostationary Synthetic Aperture RADAR Introduction} Geosynchronous and Geostationary (Geo)-SAR (GeoSAR) at the moment is only a theory because there are problems related to the target sensor distance (approx. 36000 km) that requires special attention when designing the RADAR link-budget. However, another GeoSAR main issue remains the azimuth focusing where the geostationary orbital perturbation is exploited, typically in the shape of ''eight``, and of daily period. This feature requires high integration times to obtain acceptable azimuth resolutions. In \cite{hu2011accurate}, the accurate slant range model in GeoSAR is created based on the consideration of the ''Stop-and-Go`` error assumption. All the simulation results verify the correctness and effectiveness of the proposed imaging method and resolution analysis method. 
In \cite{hobbs2014system} authors provide an overview of mission design, describing significant physical constraints. 
In \cite{sun2015inclined}, the concept and advantages of bistatic GeoSAR are firstly investigated, designing the mission dedicated at identifying a set of receiver flight parameters. In \cite{sun20132} a space-variant chirp scaling algorithm based on the range cell migration equalization and azimuth sub-band synthesis has been studied to process simulated GeoSAR data. 
In \cite{hu2016performance} authors deduced the azimuth point-spread function of GeoSAR images with the consideration of both the amplitude and phase scintillation. In \cite{hu2017three} authors focuses on 3-D deformation retrieval by GeoSAR Multiple-aperture interferometry (MAI) processing. The theoretical analysis and the experimental results suggest that centimeter-level and even millimeter-level deformation measurement accuracy could be obtained in 3-D by GeoSAR MAI processing. 
In \cite{wu2018azimuth}, the Doppler characteristics of bistatic GeoSAR and the individual contribution of the transmitter and the receiver are analyzed. In \cite{matar2019meo}, authors discuss the design tradeoffs of MEO SAR, including sensitivity and orbit selection. 
The use of these higher orbits opens the door to global coverage in one- to two-day revisit or continental/oceanic coverage with multi-daily observations, making MEO SAR very attractive for future scientific missions with specific interferometric and Polarimetric (Pol) capabilities. Finally, this research \cite{guarnieri2017atmospheric} studies the impact of the atmospheric turbulence, specifically the wet tropospheric delay, in that SAR characterized by a very long integration time, from minutes to hours, and wide swaths, such as the GeoSAR, compensating the errors with the atmospheric phase screen (APS).
\paragraph{SAR Tomography Introduction} In \cite{hu2019geosynchronous} SAR tomography (TomoSAR) techniques exploit multipass acquisitions of the same scene with slightly different view angles from the same acquisition geometry, and permits to generate full 3-D images, providing an estimation of scatterers' distribution along range, azimuth, and elevation directions. The first tomographic experiment was realized by \cite{reigber2000first}, using Full-Pol airborne SAR data. Spatial and temporal decorrelation can degrade tomographic resolution, in \cite{zhu2010tomographic} and \cite{biondi2014sar} researchers exploit the use of compressed sensing in order to perform super-resolution, while minimizing the number of interferometric tracks. In \cite{lombardini2005differential} a new interferometric mode crossing the differential SAR Interferometry and multi-temporal-baseline SAR tomography concepts, termed differential SAR tomography, is successfully proposed. Concluding, the research in \cite{tebaldini2011multibaseline} discusses about tomography in light of the experimental results from multi-temporal-baseline and polarimetric analysis of the ESA campaign BioSAR 2008.
\paragraph{MCA Introduction} MCA is a SAR investigation technique that allows to extrapolate information induced in the frequency variation generated both in range and azimuth, to synthesize classical SAR images focused at different spectral centers and bandwidths. Research \cite{bovenga2013multichromatic} uses interferometric pairs of SAR images processed at range sub-bands and explores the phase trend of each pixel as a function of the different central carrier frequencies to infer absolute optical path difference. Authors of \cite{biondi2019atmospheric} describes a complete procedure to address the challenge to estimate the atmospheric phase screen and to separate the three-phase components by exploiting only one InSAR image couple. This solution has the capability to process persistent scatterers subsidence maps potentially using only two multitemporal InSAR couples observed in any atmospheric condition. Authors of \cite{Micro_Motion_1,Micro_Motion_2} applied MCA in the azimuth direction for estimating the micro-motion of maritime vessels, and extended also for large infrastructures monitoring \cite{biondi2020monitoring}. 

To the best of our knowledge, we do not know any research report available in the literature discussing any type of OAM-based geostationary and Tomographic radar imaging technique. Here we propose a solution to implement the persistent 3D-HR RADAR observations, focusing the attention on the use of the OAM communication channel that can be considered, in the paraxial approximation orthogonal to the polarization and orthogonal to frequency and Doppler azimuth information channels. 
In practice, the observation system we propose here is characterized by four degrees of freedom and we canalize our efforts to improve the efficiency of the Interferometric OAM acquisition technique experimented for the first time in \cite{reigber2000first}. 
This technique performs a full-Pol 3D imaging of targets allowing also the scanning inside a distributed medium, according to their penetration properties and the EM carrier frequency. In this paper we increase efficiency of classical GeoSAR theory by eliminating the condition to perform GeoSAR azimuth focusing using Doppler synthesis and multi-baseline paradigm for SAR tomography, which are time consuming and mainly affected by temporal \cite{6679227}, spatial \cite{6679227} and atmospheric \cite{ferretti2001permanent,biondi2019atmospheric} decorrelation.
\paragraph{HR-RT-GeoPolInt-OAM-RADAR and HR-RT-GeoPolInt-OAM-MCA-TomoRADAR Motivations} 
This research refers to complete survey, to generate OAM, on EM waves, for real-time (RT) Geo Pol Interferometric (HR-RT-GeoPolInt-OAM-RADAR) and full-Pol tomography (HR-RT-GeoPolInt-OAM-MCA-TomoRADAR) RADAR observations. 
First we design a symmetric and asymmetric fully electronic programmable Planar Vortex Antenna (PVA) for generating ''ad-hoc`` single-mode OAM transmissions. Two-dimensional HR and RT coherent images are focused in the fast-time domain, then the tomographic slices are observed, also in the fast-time, from a unique interferometric view-point, by a stepped chirped frequency variation strategy. 
Series of HR-RT-GeoPolInt-OAM-MCA-TomoRADAR two-dimensional RADAR images are focused in the altitude direction, by fast-Fourier transforms (FFT) and synthesized by MCA. Range-azimuth-tomographic resolutions, while maintaining Rayleigh's limits, are now independent from the sensor-target distance and are related only to the induced OAM helical and the stepped-frequency bandwidths. 
In this work we will no longer mention the acronym SAR but we will go back to the origins, indicating only the RADAR acronym, this because we no longer need to synthesize any aperture in terms of Doppler band. The idea of implementing a GeoSAR, emerged to carry out the coherent observation of terrestrial sites in a persistent way. This time-delivery delay and low resolution limitations of GeoSAR, has led to migrate Earth observation satellites technology and proliferation on more practical Low-Earth-Orbits (LEO) systems. Below, is reported a list of the main problems regarding the state of the art concerning SAR sensors and where HR-RT-GeoPolInt-OAM-RADAR and HR-RT-GeoPolInt-OAM-MCA-TomoRADAR can be devoted to solve all of them:
\begin{itemize}
\item[1.] SAR acquisitions are closely related to orbital kinematics, at a few images per day and observed with different geometry, for each single space-time on Earth;
\item[2.] substantial difficulties are present in the acquisition of HR and RT SAR images using geostationary satellites \cite{8225726,7514423};
\item[3.] every SAR image is affected by layover, a perspective deformation due to  slant-looking geometry in which each target with altitude is flattened on the slant-range, azimuth plane \cite{45752};
\item[4.] the acquisition geometry must be ''side-looking``, and then nadiral SAR images cannot be acquired \cite{4323218};
\item[5.] all SAR sensors must have a certain velocity in the azimuth direction \cite{142933,158864};
\item[6.] the targets must all remain strictly stationary, in order to be observed in their true geographical position \cite{469488};
\item[7.] any SAR sensor is affected by fundamental azimuth resolution limitation, which is inversely proportional to the observed range swath \cite{6651848};
\item[8.] multi-temporal solutions are affected by heavy space-time and atmospheric decorrelation. Only few permanent scatterers can be imaged, only by long temporal series of interferometric images and after a reliable atmospheric phase screen (APS) estimation \cite{ferretti2001permanent}.
\end{itemize} 
Considering the above outlined problems, we designed an alternative method for RT single-look-complex (SLC) RADAR images formation, using geostationary orbits and somehow directly 3D and over ''short-time``, also 4D, entirely captured in the OAM-chirp fast times. We are confident that the interferometric OAM method can represent the correct solution because OAM ``communication channel'' is naturally embedded in the EM wave \cite{mignani1974dirac,tamburini2012encoding}. OAM is able to provide vorticity to EM transmissions meaning that the EM wave fronts are no longer equiphase in a plane perpendicular to the propagation direction of the EM wave and therefore the layover indetermination is automatically canceled. 
We propose the use of an interferometric solution generated by a Geo system consisting of two master (M) and slave (S) satellites, spaced one from the other by a physical baseline, in order to fully exploit the 2D characteristics of the OAM orientation vector. In fact, it seems difficult to extrapolate the 2D information from the OAM communication channel using only one antenna, with the hypothesis of using only a single OAM mode. In fact, we have preferred to remain within the single OAM transmission mode with the aim of keeping the PVAs radiation pattern as constant and stable as possible. It is clear that such a solution is of great technological interest as it would be possible to monitor the Earth's persistence, ''nigth-and-day`` and in ''any-weather``. The applications can be various, such as persistent monitoring of the territory, like rivers, car and naval traffic control and, by designing suitable fast prediction algorithms, also the global and persistent monitoring of civil air traffic, as well as real-time precise air traffic visualization on congested airports (both on the ground and in flight, when waiting for landing). Another application could also be to provide highly valuable early-warnings for dangerous tsunamis, and warning immediately the involved populations. 

The outline of this paper is the following: In section \ref{Notation} we give information about the mathematical notation used in this paper. In section \ref{Antenn_Design} we explain how to generate EM wave packets with OAM modulation, we also describe the complete design of a panel antenna. We offer a symmetrical and asymmetrical strategy, to generate OAM EM transmissions, electronically varying the transmission phase of each individual radiating element. In the section \ref{Model_and_Tomographic_Resolution} the tomographic imaging model using OAM is described and the tomographic focusing algorithm and cross-slant-range spatial resolution is also explained in detail. Section \ref{Experimental_Results} reports the experimental results on simulated data connected to subsection \ref{Performance} where the performance of a simulated OAM-MCA-TomoSAR environment is analyzed. Conclusions are finally in section \ref{Conclusions}. 
\subsection{Notation}\label{Notation}
In the following, vectors and matrices are denoted by boldface lower-case and upper-case letters, respectively. The operator (·) represents the scalar product, $\dagger$ represent the conjugate transpose operator and $\times$ is the vector product. $\nabla\cdot(\cdot)$, $\nabla\times(\cdot)$ are the gradient and the rotor operators respectively, $\partial(\cdot)$ is the partial derivative respectively, and $\prod(\cdot)$ represent rectangular windows. Finally $\cdot$ is the matrix/vector ''element-by-element`` product.
\begin{figure*}[htb!] 
	\centering
	\includegraphics[width=18cm,height=4.0cm]{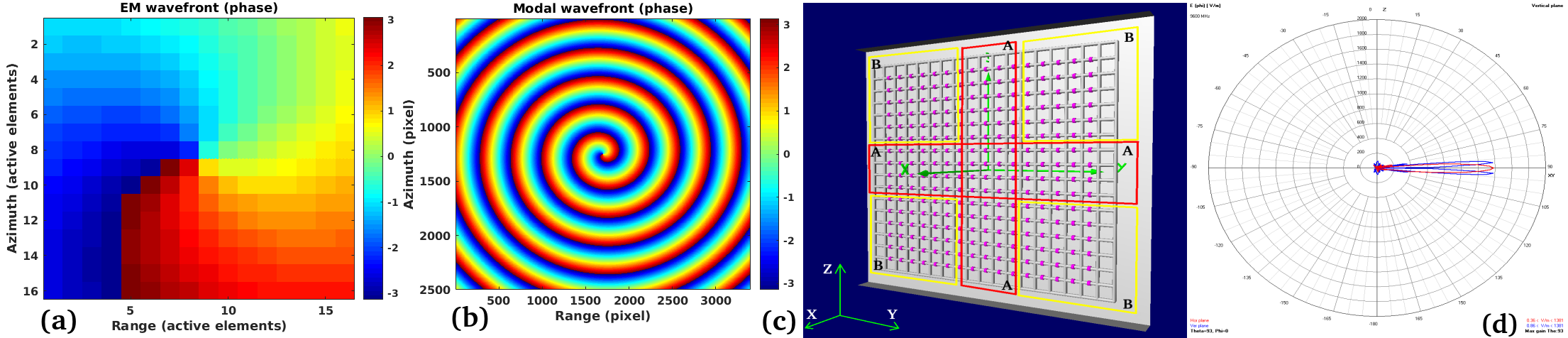}
	\caption{(a): OAM-phases allocation strategy. (b): OAM-phases Vortex propagating in space. (c): PVA structure with reflector. (d): Far-field propagation scheme Horizontal plane (red), vertical plane (blue).}
	\label{Fasi1}
\end{figure*}
\begin{figure*}[htb!] 
	\centering
	\includegraphics[width=18cm,height=4.0cm]{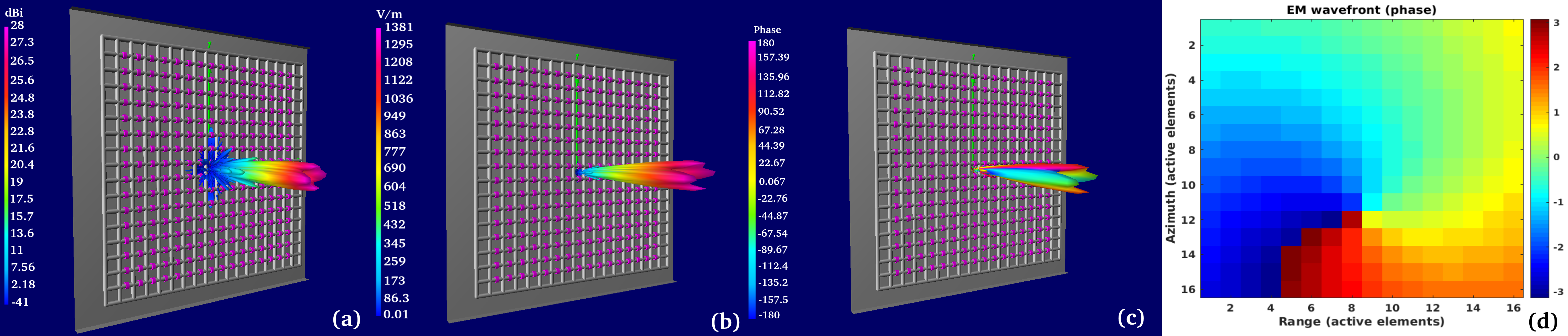}
	\caption{(a): Antenna gain 3-D pattern (b). Electric field 3-D pattern (magnitude). (c): Electric field 3-D pattern (phase). (d): OAM-asymmetric phases allocation strategy.}
	\label{Pattern_2D}
\end{figure*} 
\section{Antenna Design and OAM EM Validation}\label{Antenn_Design}
\begin{figure*}[htb!] 
	\centering
	\includegraphics[width=18cm,height=5.0cm]{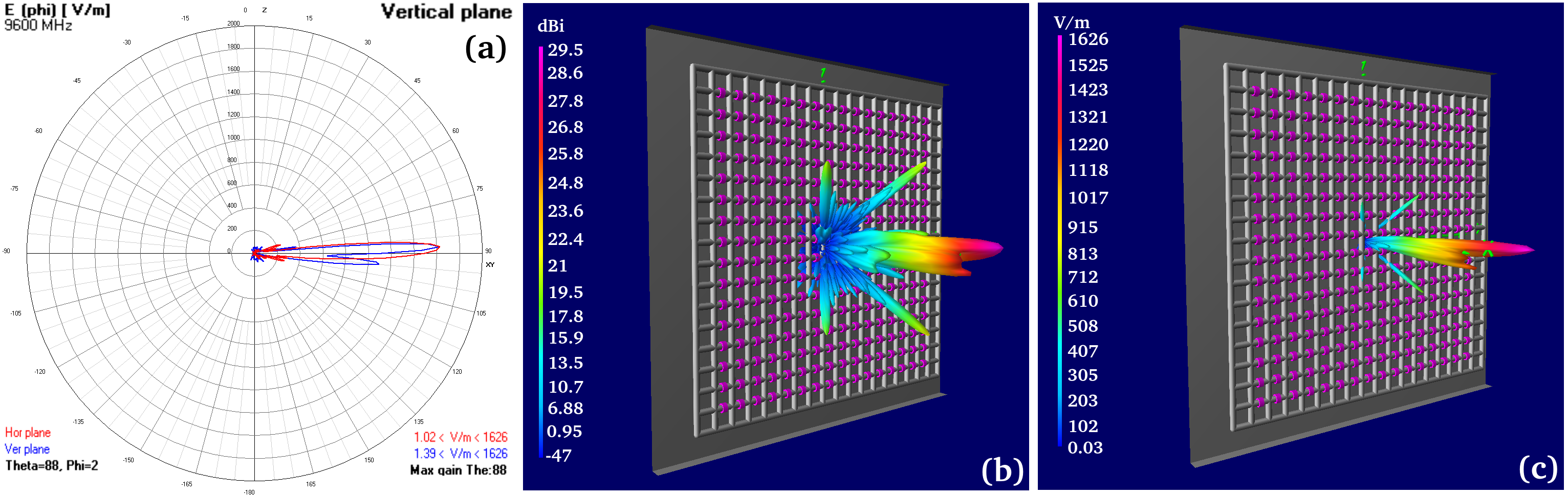}
	\caption{(a): Asymmetric Far-field propagation scheme Horizontal plane (red), vertical plane (blue). (b): Asymmetric Antenna gain 3-D pattern. (c): Asymmetric Electric field 3-D pattern (magnitude).}
	\label{Asimetric_1_4}
\end{figure*}

In this section we describe the generation of EM waves carrying OAM with our antennas. 
\paragraph{Symmetric Phase Distribution Strategy for High-Gain Vorticity Generation} 
We now describe the design of the Planar Vortex Antenna (PVA), a square-shaped panel antenna consisting of $256$ active radiating elements
and its technical details.
Moreover, we show that, by dephasing and dimensioning the 256 radiating elements, one can tune the antenna to emit an EM vortex field with a specific OAM mode and thus validate the antenna for our purposes.
These elements are positioned on a plane consisting of 16 rows and 16 columns. 
The Antenna is equipped with a perfect electric conductor reflector, where the radiating grid is slightly raised by the back metallic panel. The radiating component of the PVA consists of a grid where on each mesh we have inserted a radio-frequency generator. 
The antenna has been numerically tested, using the 4NEC2 EM and optimizer simulator software. 
The parameters of the antenna are reported in Tab. \ref{Tab_1}. 
In order to imprint a rotational phase pattern that characterizes any pure OAM mode, we start from the diagram in Figure \ref{Fasi_1} (a), showing that the phase delay strategy applied to each radiating element successfully generates an OAM wave. 
This planar allocation phases has the shape of a 2D vortex which is crucial to trigger the swirling effect of the EM wave fronts. The EM plane wave, during its propagation, assumes a phase configuration like the one represented in Figure \ref{Fasi_1} (b). 
In this configuration we have set only a single mode of vorticity. 
In Figure \ref{Fasi_1} (c) the representation of the antenna, consisting of $16 \times 16$ radiating points (represented by the purple dotted elements, is depicted. Figure \ref{Fasi_1} (d) are the  EM patterns where the red and the blue lines represent the horizontal and vertical gain patterns respectively. 
A value of $28$ dBi is reached as maximum gain for this OAM antenna. The 3D EM gain radiation pattern is depicted in Figure \ref{Pattern_2D} (a) while the electric field intensity radiation pattern is depicted in Figure \ref{Pattern_2D} (b) and its instantaneous phase-front is depicted in Figure \ref{Pattern_2D} (c). 
\paragraph{Asymmetric phase distribution strategy for high-gain vorticity generation}
As can be seen from Fig. \ref{Fasi_1} (d) and Fig. \ref{Pattern_2D} (a), the radiation pattern generated by the PVA, using a phase distribution strategy represented in Fig. \ref{Fasi_1} (a), is steered but does not seem to be totally optimized in gain.
In this paragraph we show a solution to increase the antenna gain, in order to concentrate the energy as much as possible in the selected region where there the overlapping of the two transmissions made by the M and S satellites occurs (the one existing on the common swath). The solution concerns in the asymmetry of the PVM phase distribution strategy and, to this aim, we propose the solution shown in Fig. \ref{Pattern_2D} (d) where we generated a shift in azimuth, on the PVA, of the OAM centroid. The results are very encouraging because we have obtained a considerable increase in radiation gain, up to 29.5 dBi (see Fig. \ref{Asimetric_1_4} (a)), thus obtaining an increase of $1.5$ dBi compared to the symmetrical case, (see Fig. \ref{Pattern_2D} (a)). The electric field intensity goes from a maximum of $1381$ $\frac{V}{m}$ in the symmetrical case (see Fig. \ref{Pattern_2D} (b)), to $1626$ $\frac{V}{m}$ in the asymmetrical case (see Fig. \ref{Asimetric_1_4} (b)).  Finally, in Fig. \ref{Asimetric_1_4} (c), is represented the two-dimensional diagram of the electric field energy, confirming the ability to polarize the energy on a single part of the radiation cone.
\paragraph{PVA EM Field Estimation}
Consider the following EM wave model \cite{liu2016study}:
\begin{eqnarray}\label{Eq_20}
E(\mathbf{r},\varphi)=A(\mathbf{r})\exp{\jmath\xi \phi}.
\end{eqnarray}
In Eq. \eqref{Eq_20}, the quantity $\mathbf{r}$ is the target-sensor distance vector $A(\mathbf{r})$ is the EM field amplitude without OAM, and the induced vorticity is given by the azimuthal phase dependence $\exp{i\xi \phi}$, and $\xi \in \{0,1\}$ is the is the phase modulation depth coefficient, without loosing in generality, in this paper the OAM is considered variable in a selected band at a specific wavelength, even if one can vary the vorticity across multiple wavelengths. According to Fig. \ref{Geo_Geometry_1} (c) $\{L_z,L_y\}$ are the dimensions of the antenna (width and length), expressed in pixels, and each pixel represents the individual radiating element phase delay, and are on the order of $10\lambda$, where $\lambda$ the EM wavelength so far considered, and the transmissions are set on a band with central frequency equal to 9.6 GHz. 
More details of the antenna design are listed in Tab. \ref{Tab_3}. 
The  model here proposed guarantees the analysis of a general case where $N$ is in number of the active elements in $Z$ and $M$ is the number of those present along $Y$. (In the specific case $N=M$). The $N \cdot M$ active elements are equidistantly spaced along the area in a square geometry, phased with $\delta\phi=\frac{2\pi \xi}{N} \hat{z}+\frac{2\pi \xi}{M} \hat{x}$. Considering the point-target $P(\mathbf{z},\theta,\phi)$ of Fig. \ref{Geo_Geometry_1} (c), its backscattered radiation field is:
\begin{eqnarray}\label{Eq_21}
E(\mathbf{r})=\sum_{n=0}^{N-1}\sum_{m=0}^{M-1}\frac{\exp{-\jmath \mathbf{k}|\mathbf{r}-{\mathbf{r}}_{n,m}|}}{|\mathbf{r}-{\mathbf{r}}_{n,m}|}\exp{\jmath\xi\varphi_{n,m}}.
\end{eqnarray}
\begin{table}[tb!]
	\caption{Tomographic geometry details}
	\label{Tab_3}
	\begin{center}
		\begin{tabular}{cc} \toprule
			Central frequency & 9.6 GHz\\ \midrule
			$L_y=L_z$   &10$\lambda$\\ \midrule
			reflector sides height  &0.02 m \\ \midrule
			Wires radius  &1.5 $\times 10^-3$ m \\  \midrule
			Elements hight over reflector  &0.015 m \\                   
			\bottomrule
		\end{tabular}
	\end{center}
\end{table}
In \eqref{Eq_21} $\mathbf{k}$ is the EM wave-vector, $\mathbf{r}$ and ${\mathbf{r}}_{n,m}$ are the position vectors of point $P$ and the $\{n,m\}-$th element, respectively. Considering that $|{\mathbf{r}}-{\mathbf{r}}_{n,m}|\approx {\mathbf{r}}_{e_{n,m}}-\hat{\mathbf{r}}\cdot{\mathbf{r}}_{n,m}$, and $|{\mathbf{r}}-{\mathbf{r}}_{n,m}|\approx {\mathbf{r}}_{e_{n,m}}$, and $\hat{\mathbf{r}}=\frac{{\mathbf{r}}_{e_{n,m}}}{|{\mathbf{r}}_{e_{n,m}}|}$, the normalized electric field at a generic point $P(r,\theta,\phi)$ can be given by the following approximation:
\begin{eqnarray}\label{Eq_22}
\begin{split}
E(\mathbf{{\mathbf{r}},\alpha})&\approx \frac{\exp(\jmath \mathbf{k}{\mathbf{r}}_{e_{n,m}})}{{\mathbf{r}}_{e_{n,m}}}\\
&\cdot\sum_{n=0}^{N-1}\sum_{m=0}^{M-1}\exp(-\jmath(\mathbf{k}\hat{\mathbf{r}}{\mathbf{r}}_{n,m}-\alpha \varphi_{n,m}))\\
& \approx \frac{(N\cdot M)\jmath^{-\alpha}\exp(\jmath \mathbf{k}{\mathbf{r}}_{e_{n,m}})}{{\mathbf{r}}_{e_{n,m}}}\\
& \cdot\exp{-\jmath\alpha \varphi_{n,m}}J_{\alpha}(\mathbf{k}a \sin \theta)=A_{t_{_{Tot}}}e^{\jmath 2 \varphi_{_{OAM}}}.
\end{split}
\end{eqnarray}
In \eqref{Eq_22} $J_a$ is the Bessel function of the first kind of order $\alpha=1$ because we consider a single-OAM mode and $a\approx 10\lambda$. $A_{t_{Tot}}$ is the total transmitted energy and $e^{\jmath 2 \varphi_{OAM}}$ is related to the particular OAM configuration, like the one depicted in Fig.\ref{Fasi_1} (a). For this project the radiation intensity is very different with respect to that present on a flat wavefront, with no OAM modulation. Looking at the figures \ref{Pattern_2D} (a), \ref{Pattern_2D} (a), \ref{Pattern_2D} (b), and \ref{Pattern_2D} (c), we see that there is an important null just on the main axis of the radiation pattern. In this case we call ''main lobe`` the circular crown around the main axis of the diagram. According to \eqref{Eq_22} the radiation intensity of the beam is \cite{yuan2016mode}:
\begin{eqnarray}\label{Eq_23}
U(\theta,\phi)\approx(N\cdot M)^2 J_{1}^2 (\mathbf{k} a \sin \theta).
\end{eqnarray}
The characteristic of the radiation pattern is obtained by examining
the properties of the Bessel's function. The intensity of the radiation pattern is canceled at the center of the beam and is maximum along the radial direction.
Therefore, the first lobe adjacent to the null region has the maximum gain that we call as ''mainlobe``. 
The maximum gain angle $\theta_0$ is the angle between the main axis of the radiation diagram and the angular direction of maximum gain.  It is obvious that the zero-gain region increases as the target sensor distance increases. 
Of course, it is not recommended to observe targets located in the null area. 
Initially the antenna model was derived from a disc antenna model, but we chose this design of PVA with a squared or rectangular shape, because it gives less side-lobes, more spatial directivity, and the radiation pattern provides more symmetries that can be exploited for our purposes. 
More specifically, the squared shape of the antenna ensures a better distribution of the side-lobes, in fact, observing the Fig. \ref{Fasi_1} (c), such lobes are present only inside the white boxes marked with the letter A, while they are very low, so almost absent inside the yellow bordered areas, much more extended, marked with the letter B. 
\section{Imaging Model Techniques and tomographic Resolution}\label{Model_and_Tomographic_Resolution}
This section describes the SAR imaging model, more precisely the HR-RT-GeoPolInt-OAM-RADAR and HR-RT-GeoPolInt-OAM-MCA-TomoRADAR models.

\begin{figure*}[htb!] 
	\centering
	\includegraphics[width=18cm,height=3.5cm]{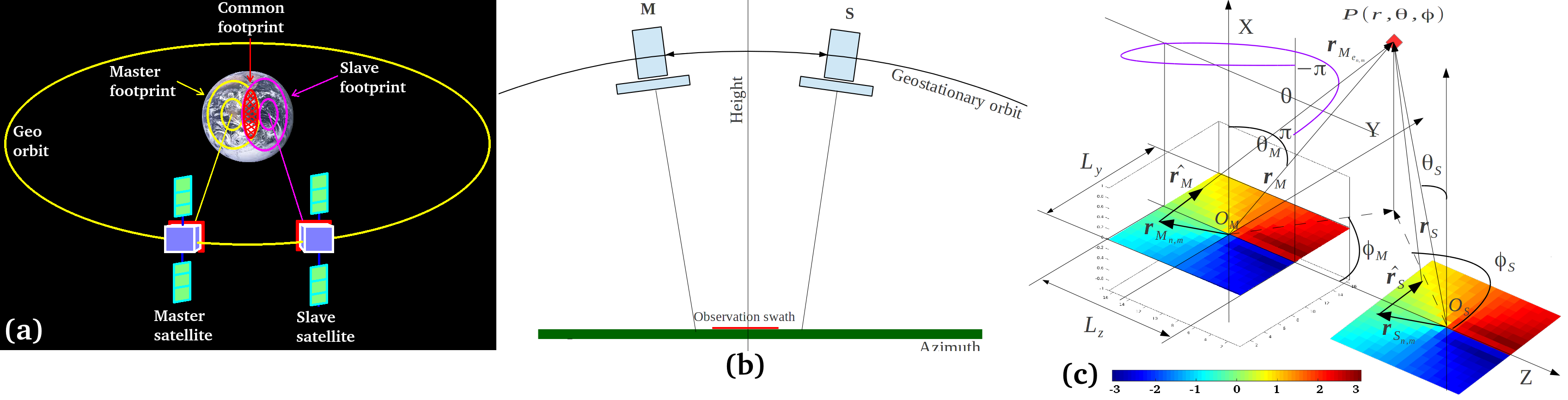}
	\caption{(a): HR-RT-GeoPolInt-OAM-MCA-TomoRADAR observation geometry: (3D view). (b): Azimuth-height view. (c): HR-RT-GeoPolInt-OAM-RADAR acquisition scheme.}
	\label{Geo_Geometry_1}
\end{figure*}

\begin{figure*}[htb!] 
	\centering
	\includegraphics[width=13cm,height=8.0cm]{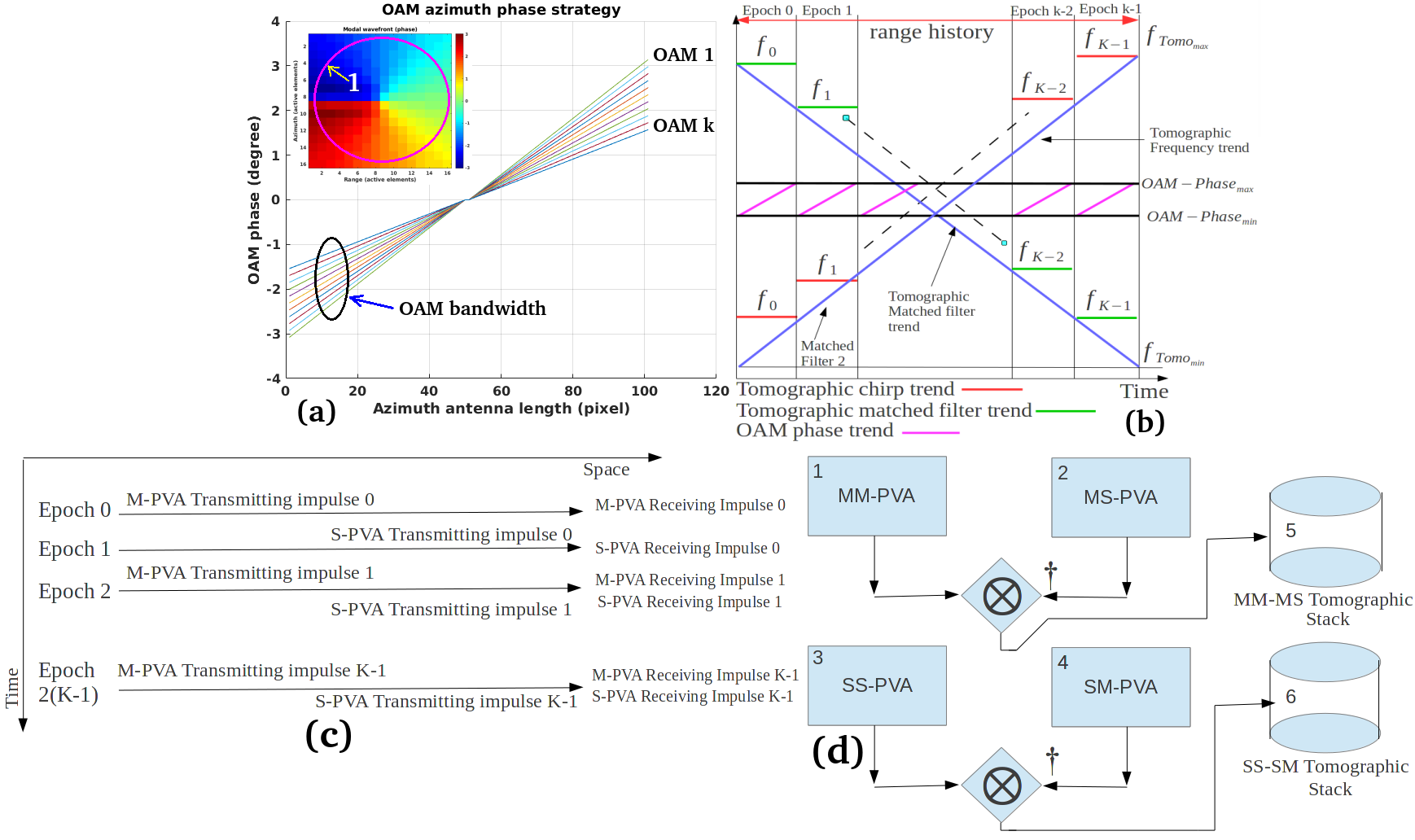}
	\caption{(a): OAM bandwidth allocation strategy. (b): Chirp frequency bandwidth and 2D OAM allocation strategy. (c): Space-time EM bursts synchronization strategy. (d): Basic computational scheme.}
	\label{OAM_Phase_Strategy_1}
\end{figure*}

\begin{figure*}[htb!] 
	\centering
	\includegraphics[width=15cm,height=5.5cm]{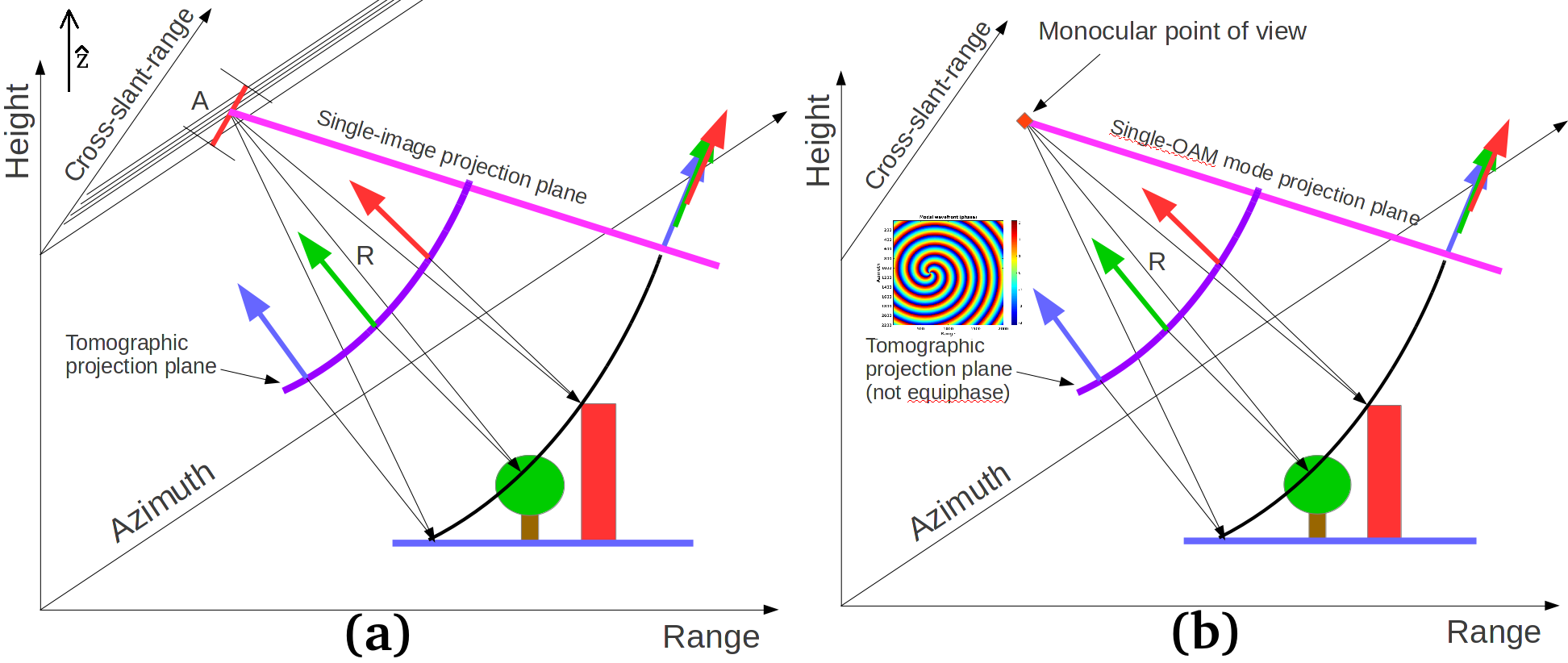}
	\caption{(a): Classical TomoSAR acquisition geometry (spatial multi-baseline). (b): HR-RT-GeoPolInt-OAM-MCA-TomoRADAR Model acquisition geometry (single view-point).}
	\label{Fig_2_QSAR}
\end{figure*}

Fig. \ref{Geo_Geometry_1} (a) shows the geometry of the geostationary acquisition system. In the figure are represented two satellites, one M and the other S, located in occupying two positions of the geostationary orbit, represented by the yellow curved line. The OAM footprint generated by the satellite M is represented in yellow, while the one represented by the satellite S is represented in red. The right active portion of the OAM footprint of M is superimposed on the active portion of the OAM footprint of S. This superimposition, visible in red, illuminates a portion of the Earth. The backscattering energy is received by one of the two satellites.  Fig. \ref{Geo_Geometry_1} (b) represents the same geometry represented in Fig. \ref{Geo_Geometry_1} (a), in the azimuth-height orientation geometry. There, are visible the two satellites M and S with their LOS, the geostationary orbit and we have approximated the section of the Earth under imaging as a plane. Finally, the SAR acquisition geometry is shown in Figures \ref{Geo_Geometry_1} (c), where each element has been reported in a $(r,\theta,\phi)$ polar coordinate reference system.
\paragraph{Received Signal Model}
Considering the measurement geometry depicted in Fig. \ref{Geo_Geometry_1} (c), the electric field component $\mathbf{E}(\alpha)$ at the point $P(\mathbf{r},\theta,\varphi)$, transmitted or by the satellite M or by the satellite S, has been formalized in \eqref{Eq_22}. We assume that the RADAR signature of $P$ is $\sigma(\mathbf{r},\theta,\varphi)$. In this model, all the elements of the PVA, used to transmit the signal, are used simultaneously to receive the backscattered echoes. According to \eqref{Eq_22} the scattered field of the $\{n,m\}-$th element and the total received signal are equal to:
\begin{eqnarray}\label{Eq_24}
\begin{split}
\mathbf{E}_{n,m}(\alpha,\mathbf{r})&=\sigma(\mathbf{r},\theta,\varphi)\mathbf{E}(\alpha,\mathbf{r})\frac{\exp\{\jmath k |\mathbf{r}-\mathbf{r}_n|\}}{|\mathbf{r}-\mathbf{r}_n|}\\
&\approx \sigma(\mathbf{r},\theta,\varphi)\mathbf{E}(\alpha,\mathbf{r})\frac{1}{r}e^{k r}e^{-\jmath k \hat{\mathbf{r}}\cdot {\mathbf{r}}_n}.
\end{split}
\end{eqnarray}
\begin{eqnarray}\label{Eq_25}
\begin{split}
	\mathbf{E}_{n,m}(\alpha,\mathbf{r})&=\sum_{n=0}^{N-1}\sum_{m=0}^{M-1}\sigma(\mathbf{r},\theta,\varphi)\mathbf{E}(\alpha,\mathbf{r})e^{i \alpha \varphi_n}\\
	&\approx \sum_{n=0}^{N-1}\sum_{m=0}^{M-1} \sigma(\mathbf{r},\theta,\varphi)\mathbf{E}(\alpha,\mathbf{r})\\
	&\cdot\frac{1}{r}e^{\jmath k r}e^{-\jmath k \hat{\mathbf{r}} \cdot {\mathbf{r}}_n}\\
	&\approx \frac{(N-1)(M-1)}{r^2} \sigma(\mathbf{r},\theta,\varphi)\\
	&\cdot J_a^2 (ka \sin \theta )e^{-\jmath \alpha \pi} e^{-\jmath 2 k r} e^{-\jmath 2 \alpha \varphi}.
\end{split}
\end{eqnarray}
In \eqref{Eq_25} the same approximation strategy of \eqref{Eq_22} has been used. The parameter $\alpha$ indicates the OAM mode, $e^{i\alpha \pi}$ is the same for all the scatterers and, of course, must be compensated together. 
The backscattered signal from $Q$ point targets, with back-scattered coefficient $\sigma_q$, for $q=\{1,\dots,Q\}$, and position is $(\mathbf{r}_m,\theta_m,\varphi_m)$, considering $\alpha=1$ one obtains:
\begin{eqnarray}\label{Eq_26}
	\mathbf{S}_r(1,\mathbf{r})=&\sum_{q=1}^{Q}\frac{\sigma_q}{r_q^2}J^2_1(k 1 \sin \theta_q)e^{\jmath 2 k r_q}e^{\jmath 2 \varphi_q}.
\end{eqnarray}
\paragraph{HR-RT-GeoPolInt-OAM-RADAR Model} In \eqref{Eq_22} we have defined the transmission model of the single satellite M or S. The back-scattering signal of $Q$ point targets is instead formalized in \eqref{Eq_26}. In this paragraph we define the structure of the interfering signal assuming that the OAM transmission is done according to the ''ping-pong`` transmission strategy, through Fig. \ref{OAM_Phase_Strategy_1} (c) and (d), that we will explain shortly. The interfering back-scattering signal of $Q$ point targets assumes the following configuration:
\begin{eqnarray}\label{Eq_27}
\begin{split}
	\mathbf{S}_r(1,\mathbf{r})_{Int}&=\mathbf{S}_r(1,\mathbf{r})_M\cdot\left(\mathbf{S}_r(1,\mathbf{r})_S\right)^\dagger\\
	&=\sum_{q=1}^{Q}\frac{\sigma_{q_M}}{r_{q_M}^2}J^2_1(k 1 \sin \theta_{q_M})e^{\jmath 2 k r_{q_M}}e^{\jmath 2 \varphi_{q_M}}\\
	&\cdot\left(\sum_{z=1}^{Q}\frac{\sigma_{z_S}}{r_{z_S}^2}J^2_1(k 1 \sin \theta_{z_S})e^{\jmath 2 k r_{z_S}}e^{\jmath 2 \varphi_{z_S}}\right)^\dagger.
	\end{split}
\end{eqnarray}
In \eqref{Eq_27} $\mathbf{S}_r(1,\mathbf{r})_{Int}$ is the interferometric signal received from a fixed interferometric view-point. If we consider $\textrm{ for } q\in\{1,\dots,Q\}$, we formalize formulas \eqref{Eq_28}, \eqref{Eq_29}, \eqref{Eq_30_Bis}, \eqref{Eq_31_Bis}, and \eqref{Eq_32_Bis}, where $\varphi_{Int_M}$ and $\varphi_{Int_S}$ are the initial phase of the M and S satellites EM signals, $\varphi_{Target_{q_M}}$ and $\varphi_{Target_{q_S}}$ are the phases of arrival concerning the EM backscattered energy by the $q-$th target and received by satellites M and S respectively. According to \eqref{Eq_28} and \eqref{Eq_29} the total received interferometric signal is equal formulas \eqref{Eq_30_Bis}, \eqref{Eq_31_Bis} and we finally present the received interferometric signal in the configuration given in \eqref{Eq_32_Bis}.
\begin{figure*}
\begin{eqnarray}\label{Eq_28}
	\mathbf{A}_{q_M}&={r_{q_M}^2}J^2_1(k 1 \sin \theta_{q_M})e^{\jmath 2 k r_{q_M}}, \mathbf{b}_{q_M}=e^{\jmath 2 \varphi_{q_M}}=e^{\jmath 2 \left(\varphi_{Init_M}+\varphi_{Target_{q_M}}\right)}.
\end{eqnarray}
\begin{eqnarray}\label{Eq_29}
	\mathbf{A}_{q_S}&={r_{q_S}^2}J^2_1(k 1 \sin \theta_{q_S})e^{\jmath 2 k r_{q_S}}, \mathbf{b}_{q_S}=e^{\jmath 2 \varphi_{q_S}}=e^{\jmath 2 \left(\varphi_{Init_S}+\varphi_{Target_{q_S}}\right)}.
\end{eqnarray}
\begin{eqnarray}\label{Eq_30_Bis}
	\mathbf{S}_{Tot_M}&=\sum_{q=0}^{Q-1}\mathbf{a}_{q_M}\mathbf{b}_{q_M}=\mathbf{a}_{Tot_M}e^{\left(\jmath 2 \varphi_{Init_M}+\jmath 2 \varphi_{Target_{Tot_M}}\right)}=\mathbf{a}_{Tot_M}e^{\left(\jmath 2 \varphi_{Scatt_M}\right)}.	
\end{eqnarray}
\begin{eqnarray}\label{Eq_31_Bis}
	\mathbf{S}_{Tot_S}&=\sum_{z=0}^{Q-1}\mathbf{a}_{z_M}\mathbf{b}_{z_M}=\mathbf{a}_{Tot_S}e^{\left(\jmath 2 \varphi_{Init_S}+\jmath 2 \varphi_{Target_{Tot_S}}\right)}\mathbf{a}_{Tot_S}e^{\left(\jmath 2 \varphi_{Scatt_S}\right)}.
\end{eqnarray} 
\begin{eqnarray}\label{Eq_32_Bis}
	&=\mathbf{S}_{Tot_M}\cdot \mathbf{S}_{Tot_S}^\dagger=\mathbf{a}_{Tot_M}e^{\left(\jmath 2 \varphi_{Scatt_M}\right)} \cdot \mathbf{a}_{Tot_S}e^{\left(-\jmath 2 \varphi_{Scatt_S}\right)}=(\mathbf{a}_{Tot_M} \mathbf{a}_{Tot_S})e^{\jmath 2\left(\varphi_{Scatt_M-Scatt_S}\right)}=\mathbf{a}_{Tot_{Int}}e^{\jmath 2 \left(\varphi_{Tot_{Int}}\right)}.
\end{eqnarray}
\end{figure*}
Considering the transmitted energy \eqref{Eq_22}, at each change of the total OAM configuration, formalized through the $e^{\jmath 2 \varphi_{OAM}}$ parameter, in terms of OAM modulation depth, (see Fig. \ref{Fasi_1} (a)), we will get a variation of the receiving OAM phase, \eqref{Eq_32_Bis} and measurable through the $e^{\jmath 2\left(\varphi_{Tot_{Int}}\right)}$ parameter. 
To form an image it is necessary to set the OAM phase variation strategy formalized in \eqref{Eq_33_Bis}, \eqref{Eq_34_Bis}, and \eqref{Eq_35_Bis}, given in the transmitted signal and reported in Eq. \eqref{Eq_22}. 
The M and S satellites are connected to each other, exchanging specific trigger signals, crucial to buildup an interferometer. 
Considering the phase variations represented in Fig. \ref{OAM_Phase_Strategy_1} (a), we have the representation of $k$ modulation depth OAM, variables from OAM$_1$ to OAM$_k$. 
The acquisition system works as follows: During the first pulse repetition time (PRT), which in this case we call it Epoch$_1=K\tau$ with $K$ the number of samples and $\tau$ is the duration of each sample. RADAR M illuminates the Earth in the configuration OAM$_1$, and at constant carrier frequency, which we suppose to be $f_c$=9.3 GHz. The S satellite illuminates the Earth by varying its OAM depth on a stepped OAM depth modulation strategy, from OAM$_1$ to OAM$_k$, also in this case the carrier frequency remains the same, equal to 9.3 GHz. Each OAM$_i$, for $i\in\{0,\dots,K-1\}$ remains constant during the sample duration $\tau$. According to Fig. \ref{OAM_Phase_Strategy_1} (a,b), Epoch$_1$ transmissions and received echoes are formalized in \eqref{Eq_33_Bis}, during Epoch$_2$ the following signal is transmitted \eqref{Eq_34_Bis} and finally, during Epoch$_K$, the \eqref{Eq_35_Bis}. According to \eqref{Eq_33_Bis}, \eqref{Eq_34_Bis}, and \eqref{Eq_35_Bis}, where the HR-RT-GeoPolInt-OAM-RADAR imaging strategy is formalized, we consider $\mathbf{e}_{T_{Int_{(i,j)}}}({\mathbf{{\mathbf{r}}},t})$ for $\{i,j\}=\{0,\dots,K-1\}$ being the generic interference generated by the transmitted signals $\mathbf{s}_{M_i}({\mathbf{{\mathbf{r}}},t})$ and $\mathbf{s}_{S_j}({\mathbf{{\mathbf{r}}},t})$ \eqref{Eq_35_1_Bis}, we define the transmission matrix \eqref{Eq_36_Bis}.
\begin{figure*}	
\begin{eqnarray}\label{Eq_33_Bis}
	\mathbf{e}_{M_1}({\mathbf{{\mathbf{r}}},t})&=\prod\left(\frac{t-\frac{K\tau}{2}}{K\tau}\right)\mathbf{a}_{t_{_{Tot_1}}} e^{\jmath 2 \varphi_{_{OAM_1}}}\cdot e^{\jmath 2\pi f_c t},\mathbf{e}^{1,{\{1,\dotsb,K-1\}}}_{S}({\mathbf{{\mathbf{r}}},t})=\sum_{z=0}^{K-1}\prod\left(\frac{t-z\frac{\tau}{2}}{\tau}\right)\mathbf{a}_{t_{_{Tot_z}}} e^{\jmath 2 \varphi_{_{OAM_z}}}\cdot e^{\jmath 2\pi f_c t}.
\end{eqnarray}
\begin{eqnarray}\label{Eq_34_Bis}
	\mathbf{e}_{M_2}({\mathbf{{\mathbf{r}}},t})&=\prod\left(\frac{t-\frac{K\tau}{2}}{K\tau}\right)\mathbf{a}_{t_{_{Tot_2}}} e^{\jmath 2 \varphi_{_{OAM_2}}}\cdot e^{\jmath 2\pi f_c t} \mathbf{e}^{2,{\{1,\dotsb,K-1\}}}_{S}({\mathbf{{\mathbf{r}}},t})&=\sum_{z=0}^{K-1}\prod\left(\frac{t-z\frac{\tau}{2}}{\tau}\right)\mathbf{a}_{t_{_{Tot_z}}} e^{\jmath 2 \varphi_{_{OAM_z}}}\cdot e^{\jmath 2\pi f_c t}.
\end{eqnarray}
\begin{eqnarray}\label{Eq_35_Bis}
	\mathbf{e}_{M_K}({\mathbf{{\mathbf{r}}},t})&=\prod\left(\frac{t-\frac{K\tau}{2}}{K\tau}\right)\mathbf{a}_{t_{_{Tot_K}}} e^{\jmath 2 \varphi_{_{OAM_K}}}\cdot e^{\jmath 2\pi f_c t}, \mathbf{e}^{K,{\{1,\dotsb,K-1\}}}_{S}({\mathbf{{\mathbf{r}}},t})=\sum_{z=0}^{K-1}\prod\left(\frac{t-z\frac{\tau}{2}}{\tau}\right)\mathbf{a}_{t_{_{Tot_z}}} e^{\jmath 2 \varphi_{_{OAM_z}}}\cdot e^{\jmath 2\pi f_c t}.
\end{eqnarray}
\begin{eqnarray}\label{Eq_35_1_Bis}
\begin{split}
	\mathbf{E}_{T_{Int_{(i,j)}}}&=\mathbf{a}_{t_{_{Tot_i}}} e^{\jmath 2 \varphi_{_{OAM_i}}}\cdot\left(\mathbf{a}_{t_{_{Tot_j}}} e^{\jmath 2 \varphi_{_{OAM_j}}}\right)^\dagger\cdot e^{\jmath 2\pi f_c t}=\mathbf{a}_{t_{_{Tot_i}}}\cdot \mathbf{a}_{t_{_{Tot_j}}} e^{\jmath 2 \left(\varphi_{_{OAM_i}}-\varphi_{_{OAM_j}}\right)}\cdot e^{\jmath 2\pi f_c t}\\
	&=\mathbf{a}_{t_{_{Tot_{(i,j)}}}}e^{\jmath 2 \varphi_{_{OAM_{(i,j)}}}}\cdot e^{\jmath 2\pi f_c t}. \textrm{ The transmission matrix is:}
	\end{split}
\end{eqnarray}
\begin{eqnarray}\label{Eq_36_Bis}
	\mathbf{E}_{T_{Int_{Tot}}}&=\begin{bmatrix}
		\mathbf{e}_{T_{Int_{(1,1)}}} & \dots & \mathbf{e}_{T_{Int_{(1,K)}}}\\
		. & \dots & .\\
		. & \dots & .\\
		. & \dots & .\\
		\mathbf{e}_{T_{Int_{(K,1)}}} & \dots & \mathbf{e}_{T_{Int_{(K,K)}}}
	\end{bmatrix} \cdot e^{\jmath 2\pi f_c t}=\begin{bmatrix}
	\mathbf{a}_{t_{_{Tot_{(1,1)}}}}e^{\jmath 2 \varphi_{_{OAM_{(1,1)}}}} & \dots & \mathbf{a}_{t_{_{Tot_{(1,K)}}}}e^{\jmath 2 \varphi_{_{OAM_{(1,K)}}}}\\
	. & \dots & .\\
	. & \dots & .\\
	. & \dots & .\\
	\mathbf{a}_{t_{_{Tot_{(K,1)}}}}e^{\jmath 2 \varphi_{_{OAM_{(K,1)}}}} & \dots & \mathbf{a}_{t_{_{Tot_{(K,K)}}}}e^{\jmath 2 \varphi_{_{OAM_{(K,K)}}}}
	\end{bmatrix}e^{\jmath 2\pi f_c t}.
\end{eqnarray}
\end{figure*}
\begin{figure*}
	\begin{eqnarray}\label{Eq_3t_Bis}
	\mathbf{S}_{R_{Int_{Tot}}}&=\begin{bmatrix}
	\mathbf{s}_{{Int}_{{Tot}_{(1,1)}}} & \dots & \mathbf{s}_{{Int}_{Tot_{(1,K)}}}\\
	. & \dots & .\\
	. & \dots & .\\
	\mathbf{s}_{Int_{{Tot}_{(K,1)}}} & \dots & \mathbf{s}_{Int_{{{Tot}_{(K,K)}}}}
	\end{bmatrix}\cdot e^{\jmath 2\pi f_c t}=	
	\end{eqnarray}
	\begin{eqnarray}\label{Eq_4t_Bis}	
	&=\begin{bmatrix}
	\mathbf{a}_{{Tot_M}_{(1,1)}}e^{\left(\jmath 2 \varphi_{{Init_M}_{(1,1)}}+\jmath 2 \varphi_{Target_{{{Int{Tot_M}}}_{(1,1)}}}\right)} & \dots & \mathbf{a}_{{Tot_M}_{(1,K)}}e^{\left(\jmath 2 \varphi_{{Init_M}_{(1,K)}}+\jmath 2 \varphi_{Target_{{{Int{Tot_M}}}_{(1,K)}}}\right)}\\
	. & \dots & .\\
	. & \dots & .\\
	\mathbf{a}_{{Tot_M}_{(K,1)}}e^{\left(\jmath 2 \varphi_{{Init_M}_{(K,1)}}+\jmath 2 \varphi_{Target_{{{Int{Tot_M}}}_{(K,1)}}}\right)} & \dots & \mathbf{a}_{{Tot_M}_{(K,K)}}e^{\left(\jmath 2 \varphi_{{Init_M}_{(K,K)}}+\jmath 2 \varphi_{Target_{{{Int{Tot_M}}}_{(K,K)}}}\right)}
	\end{bmatrix}\cdot e^{\jmath 2\pi f_c t}=	
	\end{eqnarray}
	\begin{eqnarray}\label{Eq_5t_Bis}	
	&=\begin{bmatrix}
	\mathbf{a}_{{Tot_M}_{(1,1)}}e^{\jmath 2\left(0+ \varphi_{{Target_{Tot_M}}_{(1,1)}}\right)} & \dots & \mathbf{a}_{{Tot_M}_{(1,K)}}e^{\jmath 2 \left(\frac{\lambda 2 \pi}{\xi} \left(1-K\right)+ \varphi_{{Target_{Tot_M}}_{(1,K)}}\right)}\\
	. & \dots & .\\
	. & \dots & .\\
	\mathbf{a}_{{Tot_M}_{(K,1)}}e^{\jmath 2 \left(\frac{\lambda 2 \pi}{\xi} \left(K-1\right)+ \varphi_{{Target_{Tot_M}}_{(K,1)}}\right)} & \dots & \mathbf{a}_{{Tot_M}_{(K,K)}}e^{\jmath 2 \left(0 + \varphi_{{Target_{Tot_M}}_{(K,K)}}\right)}
	\end{bmatrix}\cdot e^{\jmath 2\pi f_c t}=	
	\end{eqnarray}
	\begin{eqnarray}\label{MF_1}	
	\mathbf{S}_{MF}=\sum_{i=0}^{K-1}\sum_{j=0}^{K-1}\begin{bmatrix}
	e^{-\jmath 2\varphi_{{Target_{Tot_M}}_{(i,j)}}} & \dots & e^{-\jmath 2\varphi_{{Target_{Tot_M}}_{(i,j)}})}\\
	. & \dots & .\\
	. & \dots & .\\
	e^{-\jmath 2\varphi_{{Target_{Tot_M}}_{(i,j)}})} & \dots & e^{-\jmath 2\varphi_{{Target_{Tot_M}}_{(i,j)}}}
	\end{bmatrix}.	
	\end{eqnarray}
\end{figure*}
The variation of the OAM modulation depth assumes a configuration that is the same for both satellite M and satellite S (this is done so to obtain the same resolution in both range and azimuth direction). In \eqref{Eq_36_Bis} the term $e^{\jmath 2 \phi_{OAM_{\{i,j\}}}}$ is the generic $\{i,j\}-$th for $i,j=1,\dots,K$ stepped interferometric epoch combination of the transmitted OAM phase. In \eqref{Eq_33_Bis}, \eqref{Eq_34_Bis}, and \eqref{Eq_35_Bis}, $\mathbf{e}_{M_K} \in \mathbb{C}^{1 \times K}$, for $\{i,j\} \in \mathbb{N}$, $i,j=1,\dots,K$, and in \eqref{Eq_36_Bis} $\mathbf{E}_{T_{Int_{Tot}}} \in \mathbb{C}^{K \times K}$. Let's consider now a whole revolution of the OAM phase (the one shown in Fig. \ref{OAM_Phase_Strategy_1} (a), for which we have represented the incremental strategy from OAM$_1$ to OAM$_K$). We now formalize the OAM phase variation strategy, scaling the entire phase revolution as an angle variable from $\{0, 2\pi\}$, which, for the single mode, is exactly equal to a single physical frequency wavelength and represents the OAM helical pitch. Considering the generic OAM bandwidth existing between $\{{\xi}\lambda,\lambda\}$, the $i-$th raw and the $j-$th column of matrix \eqref{Eq_36_Bis} is the following:
\begin{eqnarray}\label{Eq_35_2_Bis}
\mathbf{E}_{T_{Int_{(i,:)}}}=&\begin{bmatrix}
	\mathbf{a}_{t_{_{Tot_i}}}\mathbf{a}_{t_{_{Tot_1}}} & e^{\jmath 2 \left(\varphi_{_{OAM_i}}-\varphi_{_{OAM_1}}\right)}\\
	.&\\
	.&\\
	\mathbf{a}_{t_{_{Tot_i}}}\mathbf{a}_{t_{_{Tot_K}}} & e^{\jmath 2 \left(\varphi_{_{OAM_i}}-\varphi_{_{OAM_1}}\right)}
\end{bmatrix} ^T e^{\jmath 2\pi f_c t}.
\end{eqnarray}
\begin{eqnarray}\label{Eq_35_2_Tris}
\mathbf{E}_{T_{Int_{(i,:)}}}=&\begin{bmatrix}
	\mathbf{a}_{t_{Tot_{(K,i)}}} & e^{\jmath \frac{2\pi\xi}{K} \left(i-1\right)}\\
	.&\\
	.&\\
	\mathbf{a}_{t_{Tot_{(K,i)}}} & e^{\jmath \frac{2\pi\xi}{K} \left(i-K\right)}
\end{bmatrix} ^T  e^{\jmath 2\pi f_c t}.\\
&\text{ for } \{i,j\} \in\{1,\dots ,K\}, K \in \mathbb{N}.\nonumber
\end{eqnarray}
\begin{eqnarray}\label{Eq_35_3_Bis}
	\mathbf{E}_{T_{Int_{(:,i)}}}=&\begin{bmatrix}
		\mathbf{a}_{t_{_{Tot_i}}}\mathbf{a}_{t_{_{Tot_1}}} & e^{\jmath 2 \left(\varphi_{_{OAM_i}}-\varphi_{_{OAM_1}}\right)}\\
		.&\\
		.&\\
		\mathbf{a}_{t_{_{Tot_i}}}\mathbf{a}_{t_{_{Tot_K}}} & e^{\jmath 2 \left(\varphi_{_{OAM_i}}-\varphi_{_{OAM_1}}\right)}
	\end{bmatrix} e^{\jmath 2\pi f_c t}.
\end{eqnarray}
\begin{eqnarray}\label{Eq_35_3_Tris}
	\mathbf{E}_{T_{Int_{(:,i)}}}=&\begin{bmatrix}
		\mathbf{a}_{t_{Tot_{(1,i)}}} & e^{\jmath \frac{2\pi\xi}{K} \left(i-1\right)}\\
		.&\\
		.&\\
		\mathbf{a}_{t_{Tot_{(K,i)}}} & e^{\jmath \frac{2\pi\xi}{K} \left(i-K\right)}
	\end{bmatrix} e^{\jmath 2\pi f_c t}.\\
	&\text{ for } \{i,j\} \in\{1,\dots ,K\},K \in \mathbb{N}.\nonumber
\end{eqnarray}
Considering \eqref{Eq_36_Bis}, the backscattering matrix has the configuration presented in \eqref{Eq_3t_Bis}, \eqref{Eq_4t_Bis} and \eqref{Eq_5t_Bis}.
\begin{figure*}[htb!] 
	\centering
	\includegraphics[width=18cm,height=4.5cm]{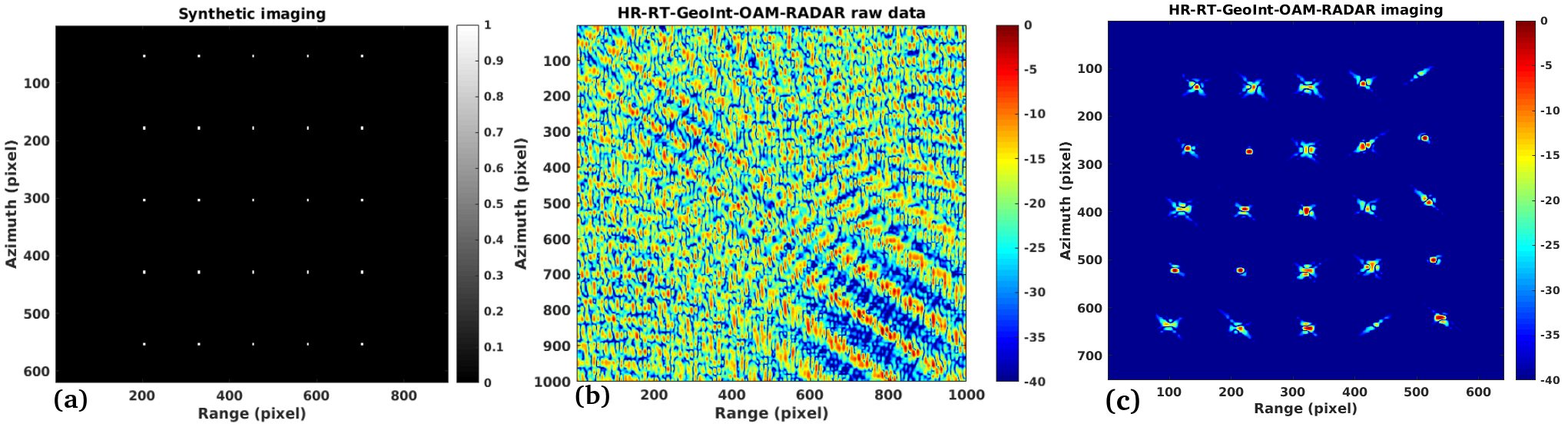}
	\caption{(a): HR-RT-GeoPolInt-OAM-RADAR synthetic targets. (b): HR-RT-GeoPolInt-OAM-RADAR raw data. (c): HR-RT-GeoPolInt-OAM-RADAR focused data.}
	\label{OAM_Radar_1}
\end{figure*}
\paragraph{HR-RT-GeoPolInt-OAM-RADAR Resolution}
Considering Fig. \ref{Geo_Geometry_1} (c), the transmitted signal \eqref{Eq_35_3_Tris}, the received signal \eqref{Eq_5t_Bis} and the adapted filter \eqref{MF_1}, given the target $P(r,\theta,\phi)$, the OAM signal, transmitted by PVA, is impressed on the ground in the shape of a spiral, so it will have a center frequency and then a wavelength. These parameters are proportional both to the modal configuration (this work considers the single mode for PVA transmission stability issues) and to the physical dimensions of the PVA, in range and azimuth dimensions. In \eqref{Eq_5t_Bis}, and \eqref{MF_1} $\mathbf{S}_{R_{Int_{Tot}}}$, and $\mathbf{S}_{MF} \in \mathbb{C}^{K\times K}$. The focused signal is:
\begin{eqnarray}\label{Foc_1}
\begin{split}
\mathbf{S}_{Foc}(x,y)&=\exp{-\jmath \frac{4\pi R_x}{L_x}}\\
&\textrm{sinc}\left(\pi B_{OAM_x}\left(x-\frac{2R_x}{c}\right) \right)\exp{-\jmath \frac{4\pi R_y}{L_y}}\\
&\textrm{sinc}\left(\pi B_{OAM_y}\left(y-\frac{2R_y}{c}\right) \right)\cdot e^{\jmath 2\pi f_c t}.
\end{split}
\end{eqnarray}
In \eqref{Foc_1} $B_{OAM_x}$ and $B_{OAM_y}$ represent the OAM band in the direction of range and azimuth respectively. In this work, for symmetry issues, we have placed $B_{OAM_x}=B_{OAM_y}=B_{OAM_{Tot}}$. The higher $B_{OAM}$ is, the more the point-spread function (PSF) has a low Rayleigh distance and therefore the range-azimuth resolution is higher. This Rayleigh distance is $\delta_{x,y}=\delta_x+\jmath\delta_y=\frac{c}{2}\left(\frac{1}{B_{OAM_x}}+\frac{1}{B_{OAM_y}}\right)=\frac{c B_{OAM_{tot}}}{B_{OAM_x}B_{OAM_y}}$ and is independent on the sensor-target distance. The computational scheme is shown in Fig. \ref{OAM_Phase_Strategy_1}. (c,d). Fig. \ref{OAM_Phase_Strategy_1} (c) is a space-time schematic representation that explains the ''ping-pong'' synchronization of the HR-RT-GeoPolInt-OAM-RADAR acquisition. during epoch 0 the M satellite transmits an OAM electromagnetic impulse, which, after the relativistic times required by the light, radiates the Earth's surface, about 36000 km (0.12s approx.) away, and emits the radar echo in all directions. A portion of the backscattered energy is received by both satellite M and S, after a second time-slot of approx. 0.12s. During Epoch 2, it is the turn of satellite S to transmit and its echoes are received simultaneously by both M and S. This procedure is repeated until Epoch $2(K-1)$. Fig. \ref{OAM_Phase_Strategy_1} (d) represents the mode on how the interferogram between generic OAM products, generated by all transmissions made by the M satellite and received by both M and S (MM-PVA, block 1 and MS-PVA block 2), and by all transmissions made by the S satellite and received, also in this case, by both M and S (SS-PVA block 3 and SM-PVA block 4). The HR-RT-GeoPolInt-OAM-MCA-TomoRADAR stacks are stored into blocks 5 and 6. The tomographic blocks are very useful because they are taken from two different geometries and therefore allow to enrich the general information content.
\paragraph{HR-RT-GeoPolInt-OAM-MCA-TomoRADAR Model} 
In this subsection we describe how to perform the tomographic focusing and we will describe in detail the concept of HR-RT-GeoPolInt-OAM-MCA-TomoRADAR resolution. For standard spatial multi-baseline TomoSAR \cite{reigber2000first} we refer to Fig.\ref{Fig_2_QSAR} (a) where the acquisition geometry is depicted. We have a cross-slant-range physical aperture $A$ and during each interferometic orbit the unique satellite observe a fully-Pol SAR data. Each SAR image is focused on the ''slant-image projection plane``. Fig. \ref{Fig_2_QSAR} (b) is the HR-RT-GeoPolInt-OAM-MCA-TomoRADAR acquisition, where we have the tomographic scheme is represented by an interferometric system immersed in 2D azimuth space. The red line in the figure represents the interferometric wavefront characterized by the HR-RT-GeoPolInt-OAM-RADAR phase ripple. As observed, satellite M transmits an incremental frequency for each time period as depicted in Fig. \ref{OAM_Phase_Strategy_1} (b), while satellite S transmits EM energy at a fixed frequency, according to the strategy that $f_{c_M}-f_{c_S}>B_{Chirp}$. where $f_{c_M}$ and $f_{c_S}$ are the central carrier and the fixed carrier frequency values of the satellites M and S respectively, and $B_{Chirp}$ is the total frequency variation of satellite M. Practically this interferometric system, functioning as a sounder, also performs tomography.
Fig. \ref{OAM_Phase_Strategy_1} (a) represents different configurations of the OAM parameter as the $\xi$ parameter changes. In this example let's suppose that the first function $OAM_1$ is equal to $\frac{1}{2}f_c$, each other increases its vorticity frequency until it reaches the value of $f_c$ where $f_c$ is the carrier frequency. Consequently the pitch of the EM wave is variable from $\frac{3}{2}\lambda$ up to $\lambda$. All the values of the space-phase characteristics have been obtained by measuring all the values around circle 1 visible in the same figure.
Here we use MCA to separate in frequency all the stepped frequencies epochs. All this is done to generate the additional sensitivity needed to focus in cross-slant range. In Fig. \ref{OAM_Phase_Strategy_1} (b) the characteristics of the adapted filter bank is represented by the green lines, visible within each epoch. Theoretically this technique has no particular limits to the tomographic resolution, strictly related to the selected frequency bandwidth, the problems are that we must have a very wide frequency band, in order to divide it into multiple zones of active chirps, so as to generate the necessary tomographic sampling in order to remain within the constraints dictated by the sampling theorem. The work done in \cite{reigber2000first} involves the coherent processing of multi-baseline interferometric SAR images, with a total baseline large enough to have an acceptable tomographic resolution, also in this case it is necessary to respect the sampling theorem and make a series of subsequent observations, made within the total tomographic aperture. Unfortunately this technique, besides having the problem of multi-temporality, the tomographic resolution is directly proportional to the total antenna length and is inversely proportional to the distance. This means that, even if we extend the antenna a lot, we have to pay attention to the spatial decorrelation, due to the aperture, but this also means that we have to perform observation within a high sampling-rate that will lead to an excessive consumption of orbital passages (for the satellite case) or avionic passages (for the aerial sensor). However the second problem is that of the distance for which it would start to be inconvenient for the satellite case. However, classic TomoSAR uses multiple Complex-valued SAR images where the phase difference between the multiple images depends on the instantaneous position of the target with respect to the sensor. Our method, is a multi-baseline generalization of \cite{bu2019synthetic}, which was done for interferometry, thus using only two OAM images. The HR-RT-GeoPolInt-OAM-MCA-TomoRADAR requires two OAM antennas, in this specific case we have designed the PVA antenna, which could transmit and receive $k-$OAM vorticity different values in a MCA fashion. After 2D imaging process using the FFT separately, the images of $k-$OAM spinning values can be expressed in the following form:
\begin{eqnarray}\label{Eq_30}
\begin{split}
\mathbf{S}_{Foc}(x,y)&=\exp{-\jmath \frac{4\pi R_x}{L_x}}\\
&\textrm{sinc}\left(\pi B_{OAM_x}\left(x-\frac{2R_x}{c}\right) \right)\exp{-\jmath \frac{4\pi R_y}{L_y}}\\
&\textrm{sinc}\left(\pi B_{OAM_y}\left(y-\frac{2R_y}{c}\right) \right)e^{\jmath 2\pi f_0 t}.
\end{split}
\end{eqnarray}
\begin{eqnarray}\label{Eq_30_bis}
\begin{split}
\mathbf{S}_{Foc}(x,y)&=\exp{-\jmath \frac{4\pi R_x}{L_x}}\\
&\textrm{sinc}\left(\pi B_{OAM_x}\left(x-\frac{2R_x}{c}\right) \right) \exp{-\jmath \frac{4\pi R_y}{L_y}}\nonumber\\
&\textrm{sinc}\left(\pi B_{OAM_y}\left(y-\frac{2R_y}{c}\right) \right)e^{\jmath 2\pi f_1 t}.
\end{split}
\end{eqnarray}
\begin{eqnarray}\label{Eq_30_bis_2}
\begin{split}
\mathbf{S}_{Foc}(x,y)&=\exp{-\jmath \frac{4\pi R_x}{L_x}}\\
&\textrm{sinc}\left(\pi B_{OAM_x}\left(x-\frac{2R_x}{c}\right) \right)\exp{-\jmath \frac{4\pi R_y}{L_y}}\nonumber\\
&\textrm{sinc}\left(\pi B_{OAM_y}\left(y-\frac{2R_y}{c}\right) \right)e^{\jmath 2\pi f_{K-1} t}.
\end{split}
\end{eqnarray}
Equations \eqref{Eq_30} consists of $k-1$ range-azimuth SAR observations, each pair of them are separated by a frequency-baseline The baseline distance is considered between the master position where the master is one of \eqref{Eq_30}, lest say $s_1(t_m,t,\xi_1)$. Let's now consider $L_1 \times L_2$ independent realizations of each range-azimuth OAM-SAR images, around the point scatterers $\{P_1, P_2, P_3\}$ represented in Fig. \ref{Geo_Geometry_1} (c), we consider the following input data vector $Y_{OAM}=\left[y(1),\dots y(k)\right] \in \mathbb{C}^{K \times 1}$:
\begin{eqnarray}\label{Input_1}
\begin{split}
{y}(1)&=\sum_{l_1=0}^{L_1}\sum_{l_2=0}^{L_2} S_{Foc}(x+l_1,y+l_2)\\
&=\exp{-\jmath \frac{4\pi R_x}{L_x}}\\
&\textrm{sinc}\left(\pi B_{OAM_x}\left((x+l_1)-\frac{2R_x}{c}\right) \right) \\
&\exp{-\jmath \frac{4\pi R_y}{L_y}}\\
&\textrm{sinc}\left(\pi B_{OAM_y}\left((y+l_2)-\frac{2R_y}{c}\right) \right)e^{\jmath 2\pi f_0 t}.
\end{split}
\end{eqnarray}
\begin{eqnarray}\label{Input_2}
\begin{split}
{y}(2)&=\sum_{l_1=0}^{L_1}\sum_{l_2=0}^{L_2} S_{Foc}(x+l_1,y+l_2)=\exp{-\jmath \frac{4\pi R_x}{L_x}}\\
&\textrm{sinc}\left(\pi B_{OAM_x}\left((x+l_1)-\frac{2R_x}{c}\right) \right)\exp{-\jmath \frac{4\pi R_y}{L_y}}\\
&\textrm{sinc}\left(\pi B_{OAM_y}\left((y+l_2)-\frac{2R_y}{c}\right) \right)e^{\jmath 2\pi f_1 t}.\nonumber
\end{split}
\end{eqnarray}
\begin{eqnarray}\label{Input_3}
\begin{split}
{y}(K-1)&=\sum_{l_1=0}^{L_1}\sum_{l_2=0}^{L_2} S_{Foc}(x+l_1,y+l_2)\\
&=\exp{-\jmath \frac{4\pi R_x}{L_x}}\textrm{sinc}\left(\pi B_{OAM_x}\left((x+l_1)-\frac{2R_x}{c}\right) \right) \\
&\exp{-\jmath \frac{4\pi R_y}{L_y}}\\
&\textrm{sinc}\left(\pi B_{OAM_y}\left((y+l_2)-\frac{2R_y}{c}\right) \right)e^{\jmath 2\pi f_{K-1} t}.\nonumber
\end{split}
\end{eqnarray}
In \eqref{Input_2} the source signal vector $S_{Foc}(x+l_1,y+l_2)$, for $\{l_1,l_2\}=0 \dots k-1$ are the complex source signal vectors given by each range-azimuth focusing process, containing the unknown complex reflection coefficients vector backscattered by $\{P_1, P_2, P_3\}$, and $\mathbf{n} \in \mathbb{C}^{K \times 1}$, represents the complex additive noise, assumed to be a Gaussian distribution with $0$ mean and variance $\sigma_n^2$, and to be white in space-time dimensions, i.e. $\mathbf{n}\sim\mathcal{N}(0,\sigma_n^2{\mathbf{I}})$ with $\mathbf{I} \ \in \mathbb{R}^{N\times N}$ is an identity matrix. The steering matrix $\mathbf{A}(\mathbf{z})=\left[\mathbf{a}(z_1),\dots,\mathbf{a}(z_F)\right]\in \mathbb{C}^{K \times F}$ contains the phase information due to frequency variation, associated to a source located at the elevation position $z$ above the reference focusing plane and is given by: $\mathbf{A}(\mathbf{z})=$
\begin{eqnarray}\label{Eq_31}
\begin{bmatrix}
1,\exp(\jmath 2\pi f_1 t z_1),\dots,\exp(\jmath 2\pi f_1 t z_0) \\
1,\exp(\jmath 2\pi f_1 t z_2),\dots,\exp(\jmath 2\pi f_1 t z_1) \\
\dots \\
1,\exp(\jmath 2\pi f_1 t z_{F-1}),\dots,\exp(\jmath 2\pi f_1 t z_{F-1})
\end{bmatrix}.
\end{eqnarray}
The standard HR-RT-GeoPolInt-OAM-MCA-TomoRADAR model is given by the following relation:
\begin{eqnarray}\label{Eq_35}
Y_{OAM}&=\mathbf{A}(\mathbf{z}) {\mathbf{h}}_{OAM-MCA-TomoSAR}(\mathbf{z}).
\end{eqnarray} 
where in \eqref{Eq_35} ${\mathbf{h}}_{OAM-MCA-TomoSAR}(\mathbf{z}) \in \mathbb{C}^{1 \times F}$, inverting \eqref{Eq_35} we finally find the following OAM-MCA-TomoSAR solution:
\begin{eqnarray}\label{Eq_34}
{\mathbf{h}}_{OAM-MCA-TomoSAR}(\mathbf{z})&=\mathbf{A}(\mathbf{z})^\dagger Y_{OAM}.
\end{eqnarray} 

\paragraph{HR-RT-GeoPolInt-OAM-MCA-TomoRADAR Resolution}
The Abbe-Rayleigh diffraction is the limit constrains of the spatial resolution of classical SAR tomography \cite{reigber2000first} calculable as:
\begin{eqnarray}\label{Eq_33}
	\delta_{csr}=\frac{\lambda R}{2A}.
\end{eqnarray}
According to multi-baseline geometry depicted in Figure \ref{Fig_2_QSAR} (a), $\lambda$ is the EM wavelength, $R$ is the target-antenna LOS distance and $A$ is the antenna aperture. In the OAM-MCA-TomoSAR the spatial resolution, which is independent from the sensor-target distance, is equal to:
\begin{eqnarray}\label{Eq_36}
	\delta_{OAM-MCA-TomoSAR}=\frac{c}{2\left(f_{K-1}-f_0\right)}=\frac{c}{2B_{Chirp}}.
\end{eqnarray}
where $c$ is the light velocity and $B_{Chirp}$ is the classical electromagnetic bandwidth variation.
\begin{figure*}[htb!] 
	\centering
	\includegraphics[width=16cm,height=4.5cm]{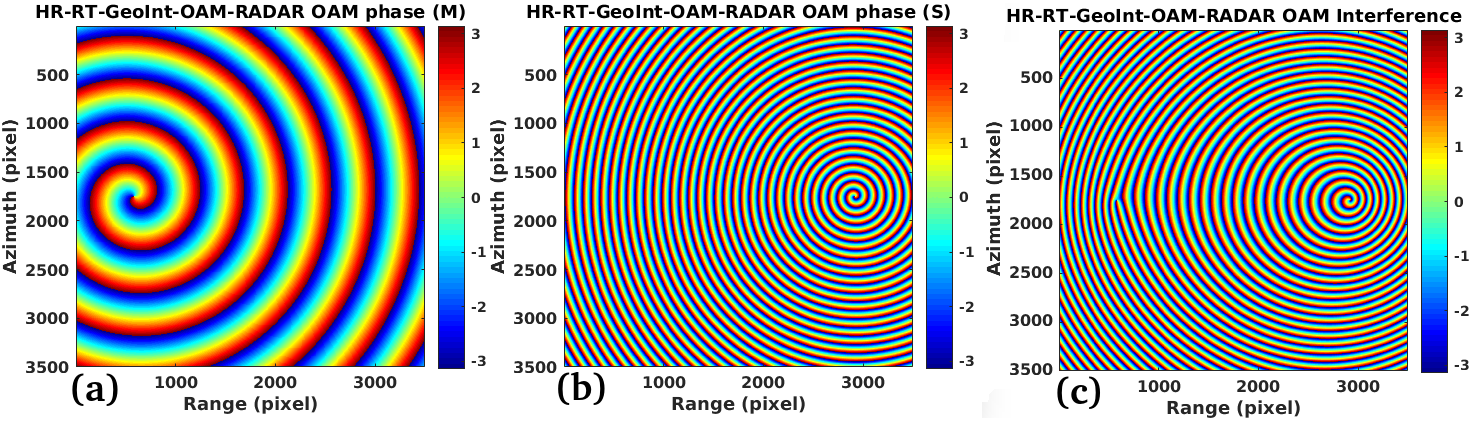}
	\caption{OAM-phases Vortexes propagating in space (a,b,c). (a): Propagation vortex of satellite M at minimum OAM frequency. (b): Propagation vortex of satellite S at maximum OAM frequency. (c): Interferometric vortex.}
	\label{Int_1_100}
\end{figure*}
\begin{figure*}[htb!] 
	\centering
	\includegraphics[width=16cm,height=4.5cm]{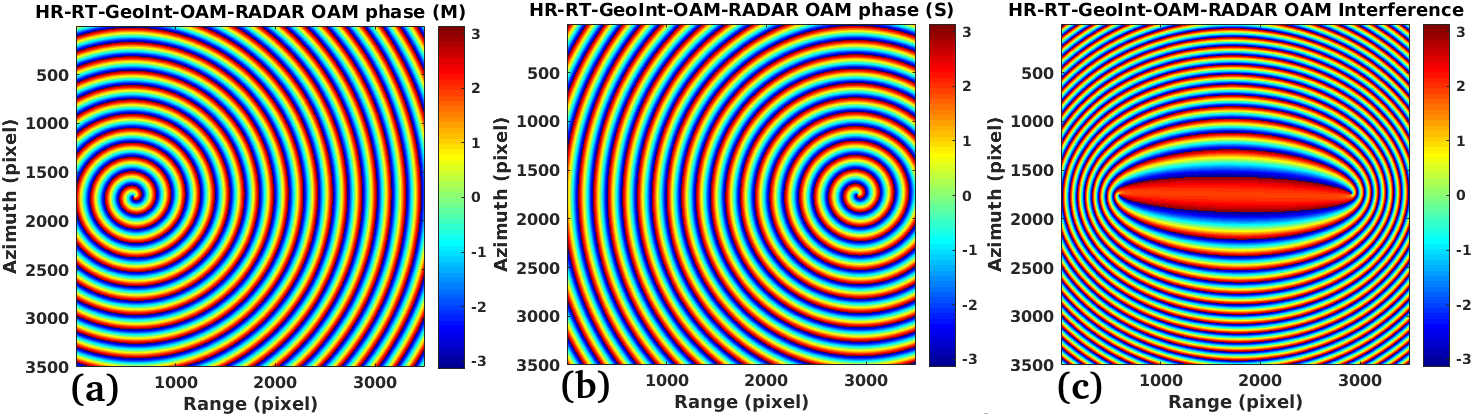}
	\caption{OAM-phases Vortexes propagating in space (a,b,c). (a): Propagation vortex of satellite M at middle OAM frequency. (b): Propagation vortex of satellite S at middle OAM frequency. (c): Interferometric vortex.}
	\label{Int_50_50}
\end{figure*}
\begin{figure*}[htb!] 
	\centering
	\includegraphics[width=16cm,height=4.5cm]{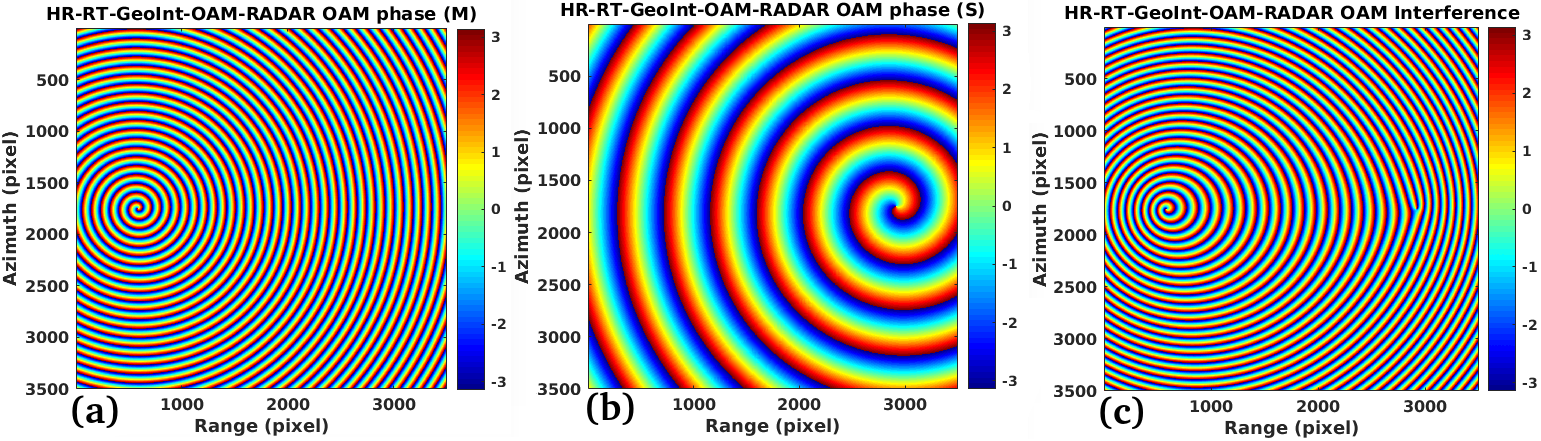}
	\caption{OAM-phases Vortexes propagating in space (a,b,c). (a): Propagation vortex of satellite M at maximum OAM frequency. (b): Propagation vortex of satellite S at minimum OAM frequency. (c): Interferometric vortex.}
	\label{Int_100_1}
\end{figure*}
\section{Results}\label{Experimental_Results}
The experimental data, obtained from numerical simulations are divided into two classes of experiments and are all performed using simulated data. The first cycle of experiments is done to validate the range-azimuth focusing belonging to the HR-RT-GeoPolInt-OAM-RADAR technique, while the second is done to validate the HR-RT-GeoPolInt-OAM-MCA-TomoRADAR technique.
\subsection{HR-RT-GeoPolInt-OAM-RADAR results} 
\begin{figure*}[htb!] 
	\centering
	\includegraphics[width=15cm,height=14cm]{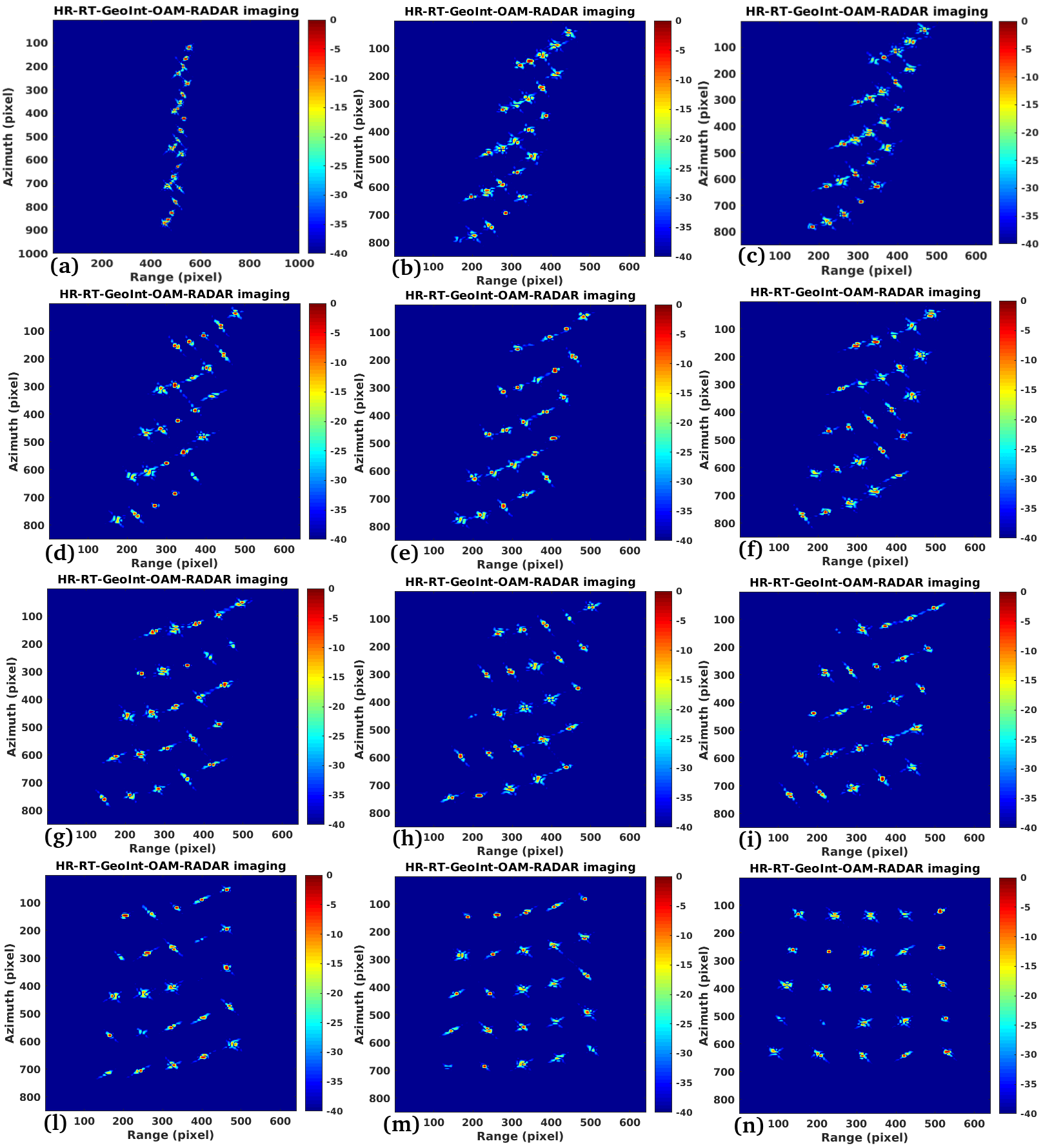}
	\caption{HR-RT-GeoPolInt-OAM-RADAR focused images at different interferometric baselines.}
	\label{Baseline_1}
\end{figure*}

\begin{figure*}[htb!] 
	\centering
	\includegraphics[width=15cm,height=14cm]{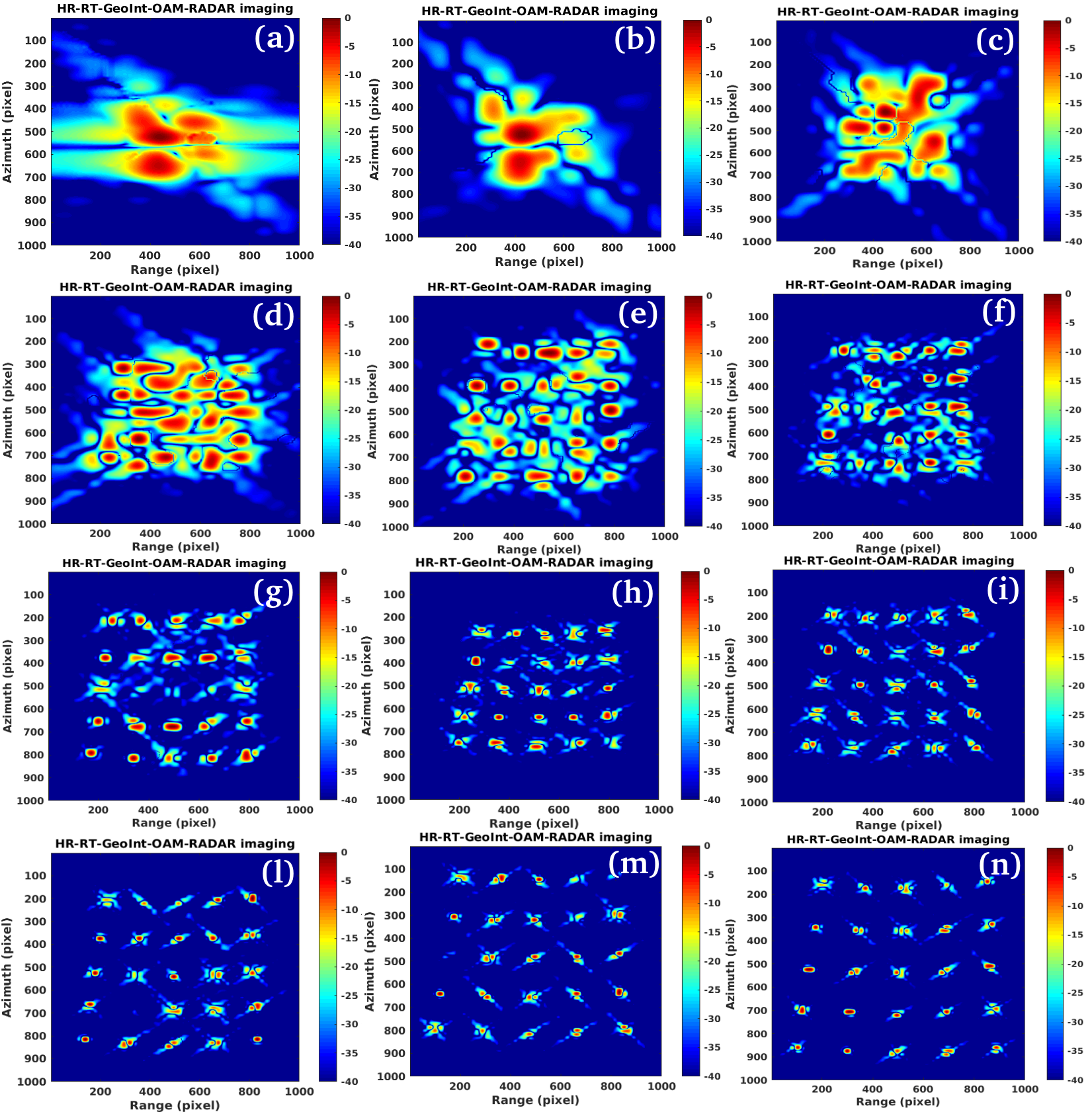}
	\caption{HR-RT-GeoPolInt-OAM-RADAR focused images at different OAM bandwidths.}
	\label{Banda_1}
\end{figure*}

The first cycle of experiments is aimed at showing the focus of a synthetic data set, namely those shown in Fig. \ref{OAM_Radar_1}. (a). The synthetic consists of 25 point scatterers arranged in an orderly manner on 5 rows and 5 columns, parallel to the range and azimuth directions. The Image shown in Fig. \ref{OAM_Radar_1}. (b) represents the raw data, that is the RADAR echoes received either from satellite M or from satellite S. Finally, the focused data is the one represented in Fig. \ref {OAM_Radar_1}. (c). Observing the result, the targets are focused in an exact way, all targets are well displayed and correctly localized in both range and azimuth direction. The results proposed in Fig. \ref{OAM_Radar_1}. (b,c) have been calculated using an ideal geometry, that is the one shown in Fig. \ref{Geo_Geometry_1} (a), (b) and (c), using a spatial baseline large enough to make the OAM interference bangs interact correctly. In this case an angular aperture between satellite M and satellite S of 25° has been set. It was found experimentally that the baseline must be there necessarily, because otherwise the interferometric effect would be lost. 
In Fig. \ref{Baseline_1} is calculated the imaging of the synthetic of Fig.\ref{OAM_Radar_1}. (a), where the M-S satellite system is sized at different space baseline aperture. This aperture is variable from a minimum of 2° to a maximum of 25°, in the following order: Fig. \ref{OAM_Radar_1}. (a) 2°, Fig. \ref{OAM_Radar_1}. (a) 5°, Fig. \ref{OAM_Radar_1}. (a) 8°, Fig. \ref{OAM_Radar_1}. (a) 12°, Fig. \ref{OAM_Radar_1}. (a) 16°, Fig. \ref{OAM_Radar_1}. (a) 18°, Fig. \ref{OAM_Radar_1}. (a) 20°, Fig. \ref{OAM_Radar_1}. (a) 22°, Fig. \ref{OAM_Radar_1}. (a) 25°. What we find is that, as the space baseline changes, there is a geometric distortion of the imaging, but this does not mean that it is not possible to acquire a RADAR image. However, the optimal baseline was found to be between 10° and 25°. If you decide to lower the baseline too much, obvious geometric distortions come out, all at the expense of spatial resolution.
Fig. \ref{Int_1_100}. (a,b,c) represent an example of OAM phases when the OAM modulation of the satellite M is at minimum helix value (\ref{Int_1_100}). (a))), while the one imposed on satellite S is at its maximum (\ref{Int_1_100}). (a,b)). The interferometric result is displayed in Fig. \ref{Int_1_100}. (c). Fig. \ref{Int_50_50}. (a,b,c) represent instead an example of OAM phases when both the OAM modulation of satellite M and the one imposed on satellite S are at equal intermediate values (\ref{Int_50_50}). (a,b)). The interferometric result is shown in Fig. \ref{Int_50_50}. (c). Fig. \ref{Int_100_1}. (a,b,c) represent an example of OAM phases when the OAM modulation of the satellite M is at maximum helix value (\ref{Int_100_1}). (a)), while the one imposed on satellite S is at its minimum (\ref{Int_100_1}). (a,b)). The interferometric result can be seen in Fig. \ref{Int_100_1}. (c). Observing the interferometric results of these three cases shows that the ground targets are illuminated, both at different spatial frequencies and with interferometric bangs located at different inclinations, i.e., orientations on the azimuth range plane, which suggests that the environment is acquired in a staring-spotlight mode, even if the sensors remain stationary. The Fig. \ref{Banda_1} represents the synthetic of Fig. \ref{OAM_Radar_1}. (a) acquired at different OAM bands, these bands range from a minimum of $0.05$ to a maximum of $0.5$
. The following list shows the band values for each sub-capture:
\begin{itemize}
	\item Fig. \ref{Banda_1} (a) $0.05\lambda)$, Fig. \ref{Banda_1} (b) $0.1\lambda)$,Fig. \ref{Banda_1} (c) $0.15\lambda)$;
	\item Fig. \ref{Banda_1} (d) $0.18\lambda)$, Fig. \ref{Banda_1} (e) $0.23\lambda)$ Fig. \ref{Banda_1} (f) $0.26\lambda)$;
	\item Fig. \ref{Banda_1} (g) $0.3\lambda)$, Fig. \ref{Banda_1} (h) $0.4\lambda)$, Fig. \ref{Banda_1} (i) $0.5\lambda)$.
\end{itemize}
Looking at the results if you understand that the more OAM band is occupied, the higher the spatial resolution, confirming the truthfulness of the \eqref{Eq_36}, where this resolution is independent of the sensor-target distance.
\subsection{HR-RT-GeoPolInt-OAM-MCA-TomoRADAR results}
We generated three scenarios, the first consisting (case study 1) of three isolated point targets located in space. The second (case study 2) is a Pol scenario consisting of a ground plane (distributed target 1), a portion of foliage belonging to a tree (distributed target 2) and two other targets simulating a building (distributed target 3). The third scenario, described in the subsection \ref{Performance}, always consists of three-point targets distributed in altitude and we show how their tomographic imaging resolution varies with the total aperture of the filters on the frequency bandwidth. The information about all cases of study is summarized in Table \ref{Tab_1}.

\begin{figure*}[htb!] 
	\centering
	\includegraphics[width=18cm,height=4.0cm]{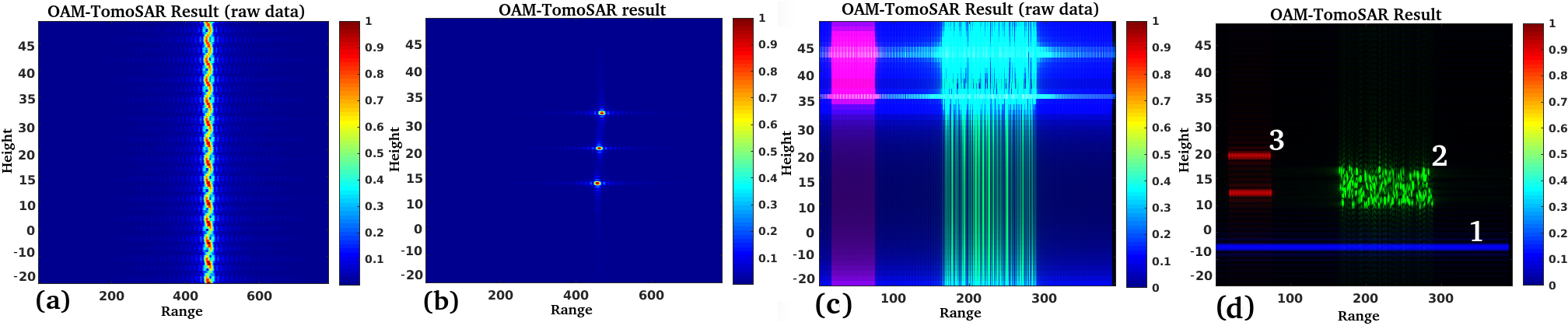}
	\caption{(a): Case study 1 raw data (magnitude). (b): Case study 1 focused data (magnitude). (c): Case study 2 raw data (magnitude) (d): Case study 2 focused data (magnitude).}
	\label{Quantum_Tomo_3}
\end{figure*}

The results are concentrated to propose a ''single-view virtual``$-$antenna tomographic focusing. We have simulated the presence of 25 filters on the frequency degree of freedom. The filters bank will surely be duplicated because the measurement is made by incorporating the possibility that the photon can be measured with different OAM, therefore with any polarization (HH, HV, VH or VV). 
Figure \ref{Quantum_Tomo_3} (a) is the OAM-MCA-TomoSAR raw data concerning the case study number one while Figure \ref{Quantum_Tomo_3} (b) is the tomographic focused data from HR-RT-GeoPolInt-OAM-MCA-TomoRADAR information. The targets are correctly focused and well distributed into space. 
For all cases of study, the geometric parameters have been designed in Tab \ref{Tab_1}
Case study number 2 is more complete. The ''synthetic`` targets environment is made up of three types of targets which, as represented in Figure \ref{Quantum_Tomo_3} (d), the distributed target number 1 is made up of flat terrain that favors the Pol representation HH+VV. The target number two simulates the foliage of a tree, modeled by small scatterers, densely collected in space. When a photon hits target number 2, with high probability there will be a change of polarization (i.e. by the SAM). Target 2 is sensitive to the HV+VH Pol channel. Target 3 is sensitive to double bounce and therefore to the HH-VV Pol channel. Figure \ref{Quantum_Tomo_3} (c) represents the case study 2 OAM raw data.
\begin{figure*}[htb!] 
	\centering
	\includegraphics[width=15cm,height=4.0cm]{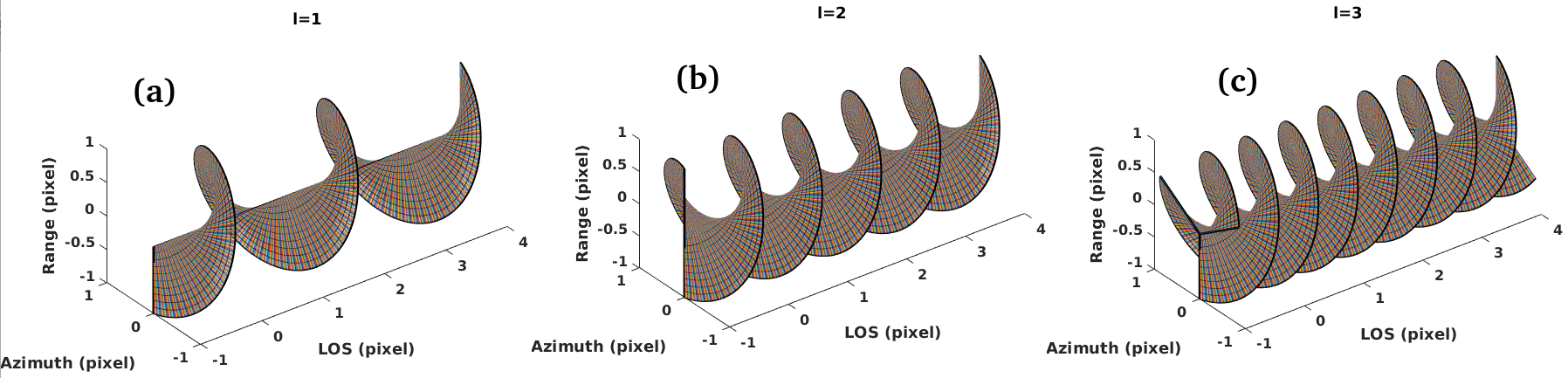}
	\caption{OAM 3-D visualization signals (multiple modes).}
	\label{L_1_Lambda}
\end{figure*}
\begin{figure*}[htb!] 
	\centering
	\includegraphics[width=15cm,height=4.0cm]{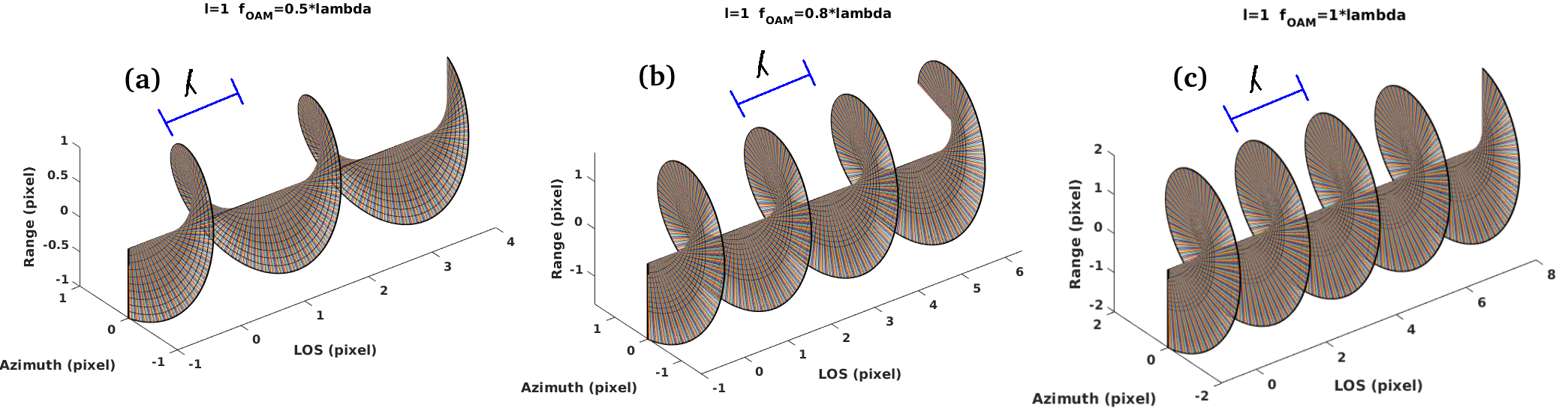}
	\caption{OAM 3-D visualization signals (single mode and OAM stepped-pitch diversity).}
	\label{f_0_5_Lambda}
\end{figure*}
\begin{figure*}[tb!]
	\centering
	\includegraphics[width=13.5cm,height=19.0cm]{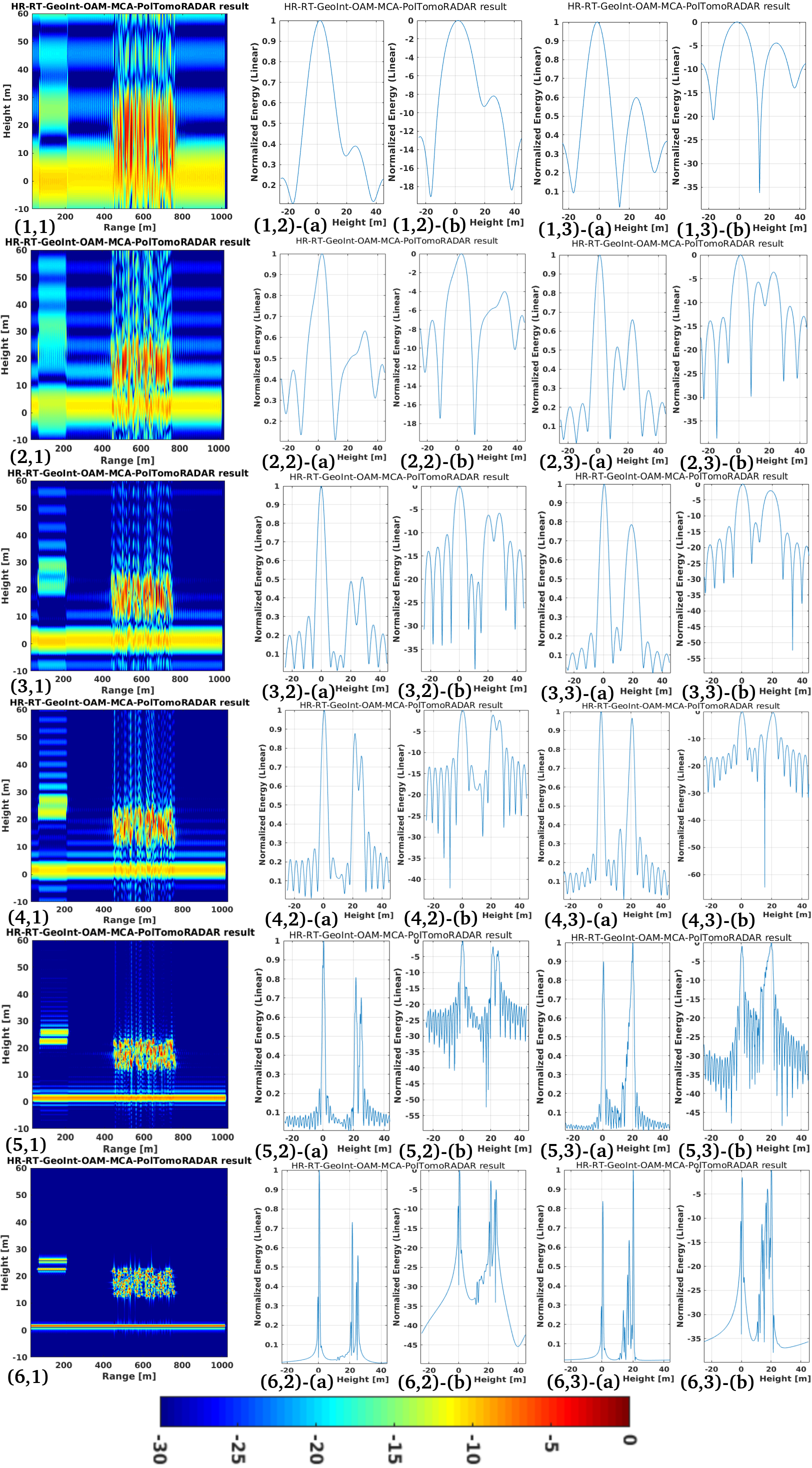}
	\caption{HR-RT-GeoPolInt-OAM-MCA-TomoRADAR Model focused data at different frequencies bandwidths.}\label{Tomo_7}
\end{figure*}
\subsection{Performance}\label{Performance}
In this subsection, the performance of the HR-RT-GeoPolInt-OAM-MCA-TomoRADAR system that observes three-point targets positioned on the same slant-range is analyzed. Simulated data are generated according to the acquisition geometry reported in \ref{Geo_Geometry_1}, 9.6 GHz center frequency, the range resolution of 0.5m. According to Tab. \ref{Tab_1} and Tab. \ref{Tab_3} we transmit EM bursts modulated with a stepped frequency stepped chirp evolution having a bandwidth varying between $f_c-250MHz$ and $f_c+250 MHz$, so the stepped frequency chirp is $B_{Chirp}=500 MHz$. At this point the theoretical tomographic resolution is $\delta_{HR-RT-GeoPolInt-OAM-MCA-TomoRADAR}=\frac{c}{2*B_{Chirp}}\approx$30 cm, independent from the LOS distance.
According to Figure \ref{Tomo_7} we have tested the same environment described in the case study two, where, the frequency bandwidth increases according to the stepped variation as described in Tab. \ref{Tab_2}, the spatial resolution increases and therefore the possibility to discriminate separately the altitude targets, and increasing the frequency bandwidth, the targets are better and better discriminated as separately.
\begin{table}[tb!]
	\caption{Tomographic geometry details}
	\label{Tab_1}
	\begin{center}
		\begin{tabular}{cccc} \toprule
			Height & Case 1 & Case 2 & Case 3 \\ \midrule
			Frequency  &9.6 GHz& 9.6 GHz & 9.6 GHz \\ \midrule
			Number of filters  &25&25&25\\ \midrule
			$\theta_{25}$  &45$^\circ$& 45$^\circ$ & 45$^\circ$  \\ \midrule
			h  &36000 km& 36000 km & 36000 km  \\  \midrule
			$B_{OAM}$  &0.3 $\frac{f_c}{\lambda}$& 0.3 $\frac{f_c}{\lambda}$ & $0-0.5*\frac{f_c}{\lambda}$   \\ \midrule
			\bottomrule
		\end{tabular}
	\end{center}
\end{table}
\begin{table}[tb!]
	\caption{Tomographic geometry details}
	\label{Tab_2}
	\begin{center}
		\begin{tabular}{cc} \toprule
				Picture number & Frequency bandwidth \\ \midrule
			1,1, (1,2; 1,3)-(a,b)  &20 MHz\\ \midrule
			2,1, (2,2; 2,3)-(a,b)  &60 MHz\\ \midrule
			3,1, (3,2; 3,3)-(a,b)  &150 MHz\\ \midrule
			4,1, (4,2; 4,3)-(a,b)  &250 MHz\\ \midrule
			5,1, (5,2; 5,3)-(a,b)  &400 MHz\\ \midrule
			6,1, (6,2; 6,3)-(a,b)  &500 MHz\\ \midrule
			\bottomrule
		\end{tabular}
	\end{center}
\end{table}
\section{Conclusions}\label{Conclusions}
In this research we designed a new paradigm for HR-RT-GeoPolInt-OAM-RADAR, and HR-RT-GeoPolInt-OAM-MCA-TomoRADAR. We proposed an alternative to the use of the Doppler channel for azimuth imaging, relying on the OAM and SAM degrees of freedom of the EM wave. 
The OAM interferometry communication channel, generated by two fixed sources separated by a spatial baseline is used for range-azimuth synthesis and the frequency channel for solving the altitude dimension. At the end, the SAM information of the EM is used to synthesize full-Pol RADAR images, with technological redundancy. We designed a planar vortex antenna, tailored for Geo applications where the imaging system transmits ''ad-hoc`` structured wave packets using an incremental stepped chirp strategy, and having single-mode OAM linearly incremented modulation. We assigned the resolutions of each dimension to three bands.  
The radial and tangential components received from the  HR-RT-GeoPolInt-OAM communication channel backscattered signals, are used to focus, through fast-Fourier transform (FFT), a range-azimuth image belonging to a single epoch at constant frequency. Each OAM fast-time RADAR image was separated in frequency by MCA. This procedure was repeated for all epochs of the entire stepped-frequency chirp. Once each two-dimensional image was synthesized, they where coregistered, and the HR-RT-GeoPolInt-OAM-MCA-TomoRADAR slices are focused in altitude, again through FFT. Range-azimuth and tomographic resolutions are dependent on the OAM and of the stepped frequency chirp bandwidths. 
We estimated from our numerical simulations to obtain, in the ideal case, a two-dimensional frame about every 10 seconds and a complete tomographic product every 2 minutes. The ''stop-and-go`` approximation is eliminated, so all targets, even those in motion, can be displayed correctly, without any delocalization or cancellation effect. 
\bibliography{manuscriptv02}{}
\bibliographystyle{ieeetr}
\end{document}